\documentclass[fleqn,usenatbib]{mnras}

\usepackage{newtxtext,newtxmath}

\usepackage[T1]{fontenc}

\DeclareRobustCommand{\VAN}[3]{#2}
\let\VANthebibliography\thebibliography
\def\thebibliography{\DeclareRobustCommand{\VAN}[3]{##3}\VANthebibliography}

\usepackage{graphicx}	
\usepackage{amsmath}	

\title[]{AMPM I. –- A Targeted Search for Asteroid Mass Primordial Black Hole Microlenses.}

\author[R. Key et al.]{Renee Key,$^{1,2}$\thanks{E-mail: rkey@swin.edu.au}
Edward N. Taylor,$^{1}$
Ken C. Freeman,$^{3}$
Jeremy Mould,$^{1,2}$
Abhijit Saha,$^{4}$
Anais M{\"o}ller,$^{1}$
\newauthor
Timothy M. C. Abbott,$^{5}$
and Alan R. Duffy$^{1,2}$
\\
$^{1}$Centre for Astrophysics and Supercomputing, Swinburne University of Technology, Melbourne, 3122, VIC, Australia\\
$^{2}$ARC Centre of Excellence for Dark Matter Particle Physics\\
$^{3}$Research School of Astronomy and Astrophysics Australian National University, Canberra, 2611, ACT, Australia\\
$^{4}$NSF NOIRLab National Optical-Infrared Astronomy Research Laboratory, 
950 North Cherry Ave., Tucson, 85719, AZ, USA\\
$^{5}$Cerro Tololo Inter-American Observatory, Casilla 603, La Serena, Chile}
\date{Accepted XXX. Received YYY; in original form ZZZ}

\pubyear{\the\year{}}

\begin{document}
\label{firstpage}
\pagerange{\pageref{firstpage}--\pageref{lastpage}}
\maketitle

\begin{abstract}
Gravitational microlensing is a powerful technique for constraining the abundance of dark matter in asteroid mass to supermassive primordial black holes, at masses of $-11 \lesssim \log M/\mathrm{M}_\odot \lesssim 5$ \citep{DeRocco_2024, Lu_2019, Carr_2024}. In this work, we introduce a new high-cadence stellar microlensing survey in the Large Magellanic Cloud, AMPM. The primary goal of AMPM is to place constraints in the asteroid-to-planetary-mass regime of primordial black hole dark matter. We present the five nights of survey data, the microlensing detection pipeline, and the microlensing efficiency of AMPM. We explore the impact of the stellar distribution in the Large Magellanic Cloud on the microlensing detection efficiency and conduct a detailed analysis of second-order microlensing effects and the impact on the primordial black hole dark matter constraints. Our findings indicate that these second-order effects shift the maximum sensitivity of AMPM toward the lunar-mass black hole regime at $10^{-8} - 10^{-6} \, M_{\odot}$. From the five nights of data, we detect a single microlensing candidate and find that AMPM can constrain at the $95\%$ C.L up to 30\% of the Galactic primordial black hole dark matter distribution. \end{abstract}

\begin{keywords}
gravitational lensing: micro -- galaxies: Magellanic Clouds -- dark matter
\end{keywords}



\section{Introduction}
As a subfield of transient astronomy, microlensing is among the only tools available for discovering distant, cold, and compact objects \citep{Paczyski_1996}. Among such objects include rocky or gas planets on wide orbits \citep{Zang_2025, Poleski_2021, Han_2005} or free-floating \citep{Dong_2026, Han_2004}, and primordial black holes, see \citet{Carr_2025} for a review. A microlensing experiment involves monitoring a field of background sources to detect microlenses in the intervening volume of space, so that the number of events observed depends both on the distribution of lens masses and the number of observed source stars. The duration of each event is determined by the lens mass and the relative motion of the source and lens \citep{Paczynski_1986, WittMao_1994}. Where microlenses with the mass of Jupiter can be found with day cadence or greater \citep{Ju_2022, Sumi_2011}, Earth-mass microlensing events require observations at a cadence of 30 minutes or less \citep{Mroz_2020_terrestrial, Beaulieu_2006}.

As such, the survey's sensitivity to a minimum lens mass depends primarily on the observational cadence.  Improvements in detector efficiency and read-out times have pushed the frontiers of fast-cadence microlensing. From the 1990s, the usual cadence of microlensing surveys was in the order of a day (e.g MACHO and EROS \citep{Alcock_2000, Tisserand_2007}, and has gradually reduced to 15-30 minutes with OGLE, MOA and Kepler \citep{Mroz_OGLEIV2024, Nunota_2025, Griest_2014}. Most recently microlensing detections have been made with two-minute cadence from TESS \citep{Kunimoto_2025} and Subaru-HSC \citep{Niikura_2019}. 

One of the primary motivations for these microlensing surveys was to identify compact dark matter candidates and place limits on the fraction of such objects as the totality of dark matter \citep{Carr_1990, Alcock_2000, Alcock_2001}. Since the objects of the dark matter search were hypothesised to live in the halo around the Galaxy, these surveys targeted star fields in nearby galaxies to maximise the probability of a halo microlensing event. The detection of exoplanets via microlensing assumes that planets are co-located with their host stars, such that the target fields are selected from the Milky Way bulge and disk \citep{Mroz_OGLEIV2024, DeRocco_2023, Niikura_OGLE2019}. The Large and Small Magellanic Clouds (LMC, SMC) and M31 are both highly studied fields for galactic dark matter surveys. 

Among the compact dark matter candidates, primordial black holes (PBHs) have become increasingly popular in the years following the LIGO-Virgo merger detections \citep{Bird_2016, Salucci_2019, Clesse_2022}. Such black holes are hypothesised to form pre-inflation, so they trace pre-inflationary physics and are also invoked as potential seeds for supermassive black hole (SMBH) formation \citep{Liu_2022, Dayal_2024, Colazo_2024}. Other dark matter compact objects include axion stars \citep{Fujikura}, fermi balls \citep{Flores_2023} and dark white dwarfs \citep{Ryan_2022}.

The dark matter microlensing experiments have ruled out cosmologically significant amounts of PBH-dark matter in between $10^{-11} < M_{PBH}/M_{\odot} < 10^{5}$ \citep{Carr_2024, Green_2026}. Towards the low-mass end of the PBH spectrum, several limits from fast-cadence observations have been established. Using data from 150,000 stars in the Kepler measurements of the Cygnus-Lyra region, \citet{Griest_2014} were able to extend the searchable range of PBHs to cover $2\times10^{-9} < M_{PBH}/M_{\odot} < 10^{-7}$. From a null detection, the strongest constraint from disallowed PBHs of mass $10^{-8} M_{\odot}$ from being more than 50\% of the dark matter. More recently, the fourth phase of the OGLE experiment, OGLE$IV$, established extremely stringent microlensing constraints based on a null result from a survey lasting 465 days and observing 35 million stars in the LMC and SMC at 16-minute cadence. Two candidate events were identified in the OGLE$IV$ survey, but both were attributed to contamination—one from a flaring star and the other from a self-lensing star. OGLE$IV$ concluded that PBHs between $10^{-8} M_{\odot}$ and $10M_{\odot}$ cannot constitute more than 10\% of dark matter, with the tightest constraint being to 1\% of the dark matter at $10^{-6} M_{\odot}$ \citep{Mroz_OGLEIV2024}. 
Returning to the question of FFP versus PBH microlens, \citet{Niikura_OGLE2019} took five years of OGLE galactic bulge observations \citep{Mroz_2017, Mroz_2019}, which were primarily dedicated to locating microlensing planets, and considered the six detected ‘ultra-short’ ($\sim 0.1$ days) microlensing events as PBHs rather than the original FFP status. The six events allowed PBH dark matter in the galactic bulge to be constrained to 1-0.1\% between masses of $10^{-6} - 10^{-4} M_{\odot}$.

The Subaru-HSC survey images M31 at a cadence of two minutes, which allowed for PBHs with masses between $10^{-11}$ and $10^{-5} M_{\odot}$ to be limited to 0.1\% of the dark matter. A single PBH candidate was found from 7 hours of observations, with a short timescale suggesting the microlens was sub-Earth mass \citep{Niikura_2019}. \citet{Smyth_2020} used a realistic distribution of source star radii in M31 to reanalyse the Subaru-HSC limits. The inclusion of a wide range of star sizes ($4-15 R_{\odot}$) reduced the sensitivity of Subaru-HSC to finding low-mass PBHs from $10^{-11}$ to $10^{-10} M_{\odot}$, and weakened the dark matter limits to 6\%.Recently, Subaru-HSC has reported 12 new PBH candidates from a combined 39.5 hours of M31 observations at the 2-minute cadence. The updated PBH limits constrain black holes above $10^{-10} M_{\odot}$ to be no more than 2\% of the dark matter. A likelihood analysis places the masses of the 12 PBH candidates to be around $10^{-8} M_{\odot}$ \citep{Sugiyama_2026}. If these candidates are indeed PBHs with mass below the Tolman–Oppenheimer–Volkoff limit of $3 M_{\odot}$, it would confirm the PBH as black holes formed from alternative, inflationary mechanisms separate from stellar collapse black holes \citep{Villanueva_Domingo_2021}, and inform the assumptions and models of inflation that are capable of producing such objects \citep{Green_2024}.

This paper presents a pilot DECam microlensing experiment, AMPM (Asteroid-Mass Primordial black hole Microlensing), designed to test the limits of fast-cadence microlensing for PBHs as dark matter. Using DECam’s wide field of view and sub-minute cadence to image the LMC, AMPM aims to locate low-mass PBH microlenses.

The structure of this paper is as follows: in Section 2, we present the relevant mathematical equations for gravitational microlensing and discuss the second-order effects that are prevalent in low-mass microlensing events. Section 3 describes the technical motivation for AMPM and the image collection and photometric reduction. In Section 4, we assess AMPM's efficiency by evaluating a realistic Galactic simulation along the AMPM sightline. We use a finite-source treatment of microlensing detection, including an analysis of the limiting PBH masses that are both theoretically and observationally detectable in AMPM.  Section 5 describes the detection method for locating fast microlensing signals in the data. We present the various flexible statistics used to evaluate microlensing candidates. The PBH microlensing rate for AMPM and the resulting dark matter constraints are presented in Section 6. We also discuss the implications and limitations of the dark matter density constraints. Lastly, we provide concluding remarks on the results from AMPM and provide an outlook for the future of asteroid-mass PBH dark matter.

\section{Gravitational Microlensing}\label{sec:maths}
\begin{figure}
\centering
\includegraphics[trim={0 0 0 0},clip,width=1.\columnwidth]{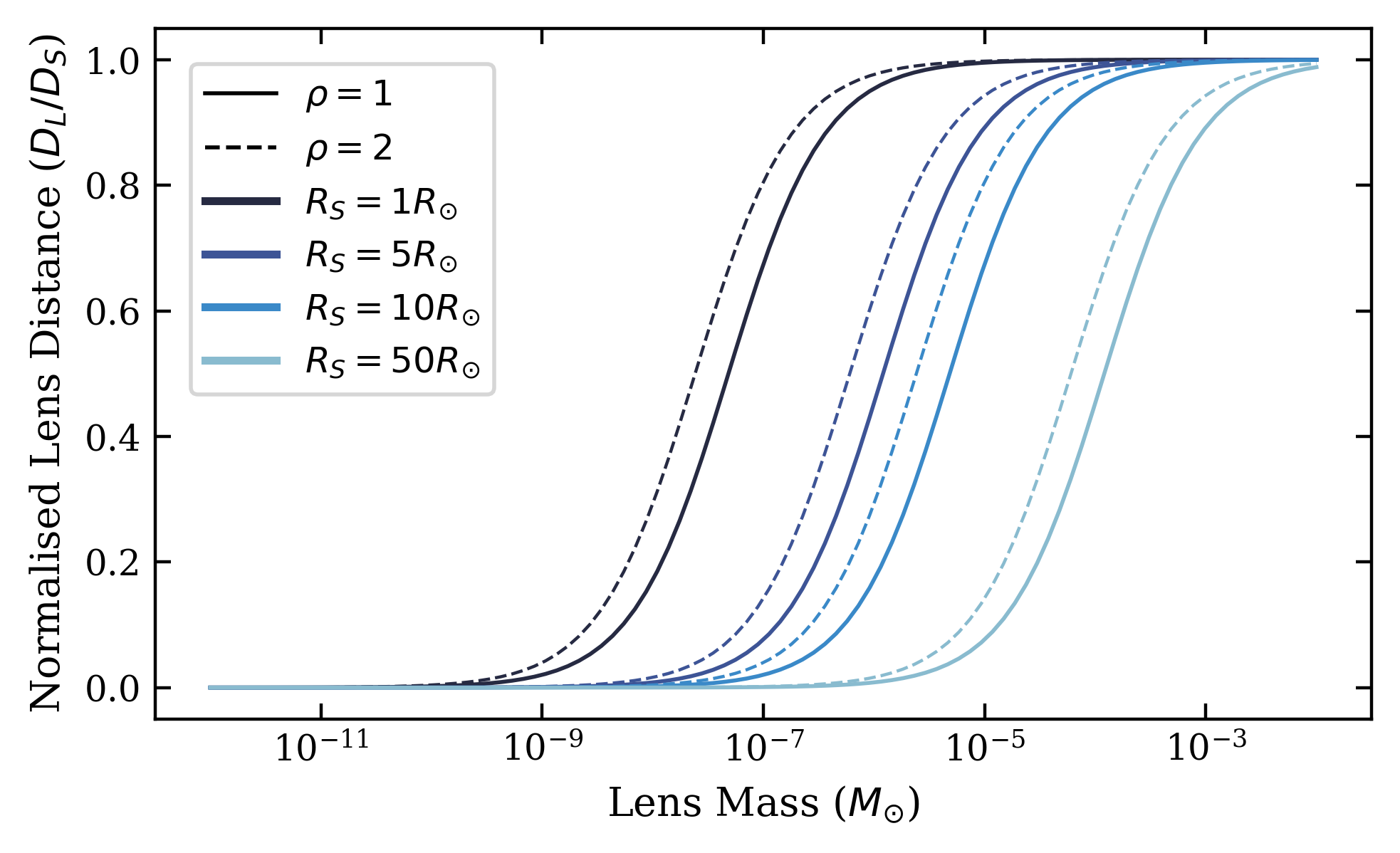}
\caption{Examples of the maximum distance, normalised by the source distance, to a lens given two $A_{max}$ thresholds defined by $\rho = 1$ is the solid line and $\rho = 2$ as the dashed line. An increasing source radius for a constant $A_{max}$ threshold brings the lens closer to the observer. Similarly, an increase in the $A_{max}$ value from an increase in $\rho$ further draws the lens towards the observer.}\label{fig:xmaxdistances}
\end{figure}

Gravitational microlensing is the unresolved strong lensing produced by an intervening massive, compact object moving between an observer on Earth, and a distant source star. As the source is gravitationally lensed two images are produced. The lens is small compared to the source star, the images are separated only by microarcseconds \citep{Schneider_1992, WittMao_1994}, with radial projection from the source given by the Einstein ring radius \citep{NO_1984}.
\begin{equation}
    R_{E} = \sqrt{\frac{4 \, G \, M \, D_{L} (D_{S} - D_{L})}{c^{2} \, D_{S}}}\, .
\label{eq:rE}
\end{equation}
Here, $M$ is the mass of the microlens, $D_{L}$ is the distance to the lens from the observer, $D_{L}$ is the distance to the source star from the observer, and $c$ and $G$ are the speed of light and gravitational constant.
The combination of the two lensed images co-adds to the source flux and magnifies the star's brightness. The compact microlens moves with some trajectory in time, $u(t)$, such that the microlensing amplification is a transient event. The amplification of a microlensing event, assuming the source is an infinitesimal point of light, is given by \citep{VO_1983, Paczynski_1986},
\begin{equation}
    A(u)_{PS} = \frac{u^{2} + 2}{u \sqrt{u^{2} + 4}} \, ,
\label{eq:Aps}
\end{equation}
where $u(t)= \sqrt{u_{0}^{2} + (\frac{t - t_{0}}{t_{E}})^{2}} $, $t_{0}$ is the time at peak amplification, and $u_{0}$ is the impact parameter, or minimum separation between lens and source at $t_{0}$, normalised by the Einstein radius. 
\newline\newline
The timescale of the event, $t_{E}$, is the time a lens moving at a transverse velocity ($v_{\perp}$) takes to travel the Einstein radius ($R_{E}$); which, in turn, depends on the mass of the lens ($M$) and the distance to the lens and source ($D_{L}$, $D_{S}$),
\begin{equation}
    t_{E} = \frac{R_{E}}{v_{\perp}}\, ,
\label{eq:tE}
\end{equation}
The timescale of the microlensing amplification depends on the mass of the lens; lower-mass microlenses produce faster signals \citep{Paczyski_1996}.
\newline\newline
The microlensing amplification given by Equation \ref{eq:Aps} models the source and microlens system as infinitely small, point-like objects. Most classes of astrophysical compact objects will satisfy a point-lens assumption, which only requires that the physical extent of the lens is smaller than the Einstein radius $R_{E}$. For a PBH microlens, a point-like assumption is always sufficient, as the Einstein ring radius is by definition larger than the Schwarzschild radius of the black hole. The physical size of the source, however, strongly affects microlensing amplification. As the mass of the microlens decreases, the separation between the lensed images accordingly reduces. For stars close to the observer (such as those in the Large Magellanic Cloud), and/or for very low-mass microlenses, the separation between the images becomes comparable to the angular size of the source. A point-like source assumption is no longer valid \citep{Smyth_2020, Griest_2011}. The parameter $\rho$ quantifies the ratio between the angular Einstein radius of the lens ($\theta_E$) and the angular radius of the star ($\theta_{S}$) \citep{WittMao_1994, Griest_1991}:
\begin{equation}
    \rho = \frac{\theta_{S}}{\theta_{E}} = \frac{R_{S}\,D_{L}}{R_{E}\,D_{S}} \, ,
\end{equation}
where $R_{S}$ refers to the physical radius of the source. A finite-source treatment in microlensing is incorporated by integrating the amplification curve ($A_{PS}$) over the source's 2D disc \citep{Lee_2009}. 
\begin{equation}
\label{eq:AFS}
A_{FS}(u, \rho) = \frac{2}{\pi\,\rho}\int_{0}^{\pi}\int_{u_{1}}^{u_{2}} A_{PS}(u')\, u' \, du' \, d\theta \, .
\end{equation}
The finite-source treatment of microlensing both dampens and broadens the amplification curve. Naturally, as the finite source amplification depends on $\rho$, the effect is more pronounced at smaller lens masses and for larger stars. The maximum amplification of the FS$-$PL event as $u_{0}$ tends towards zero is bounded as \citep{Paczyski_1996, Cieplak_2013},
\begin{equation}
    A_{max} = \frac{\sqrt{4 + \rho^{2}}}{\rho}
\label{eq:amax}
\end{equation}
Given that $\theta_{E}$ depends on the microlens' distance and mass, there is some maximum amplification limit $A_{max}$ for any lensing configuration. A distance limit ($D_{max}$) exists between the observer and source from where the maximum event amplification will always fall below $A_{max}$. Rearranging equation \ref{eq:amax} yields,
\begin{equation}
    D_{max} = \frac{2\,R_{E}\,D_{S}}{R_{S}\sqrt{1 + A_{max}^2}} \, .
\label{eq:dmax}
\end{equation}
Figure \ref{fig:xmaxdistances} shows the boundary of $D_{max}$ given the microlens mass for a range of LMC stellar radii. As the source size increases, the maximum distance threshold approaches the observer. To produce a successfully significant microlensing event above the $A_{max}$ threshold given by $\rho =1 $, a $10^{-9}M_{\odot}$ lens on a $1\,R_{\odot}$ star cannot be more than $0.02$ of the distance to the source. For the same lens on a $10\,R_{\odot}$ star, the lens must be closer by a factor of one hundred.

The optical depth quantifies the probability of an individual star passing through the
lensing cross-section of a microlens distributed between the observer and star with some varying density distribution $\varrho(DL)$. Finite source effects are  incorporated into the optical depth by limiting the line-of-sight to the source star to $D_{max}$.
\begin{equation}
    \tau =\frac{4\pi G u_{T}^{2}}{c^{2}}\int^{D_{max}}_{0}\varrho({D_{L}})\frac{D_{L}(D_{S} - D_{L})}{D_{S}}dD_{L}
\label{eq:tau}
\end{equation}
Similar to the the $D_{max}$ threshold, the $u_{T}$ factor represents the value for $u$ defined by a set amplification threshold $A_{T}$. For the remainder of this paper, we refer to the point-source point-lens approximation as PS-PL, and the finite-source point-lens treatment as FS-PL.


\section{LMC Observations and Photometry} \label{sec:obs}
\subsection{The AMPM Observations}
\begin{figure}[h]
\centering
\includegraphics[trim={11cm 1cm 8cm 0cm},clip,width=1.\columnwidth]{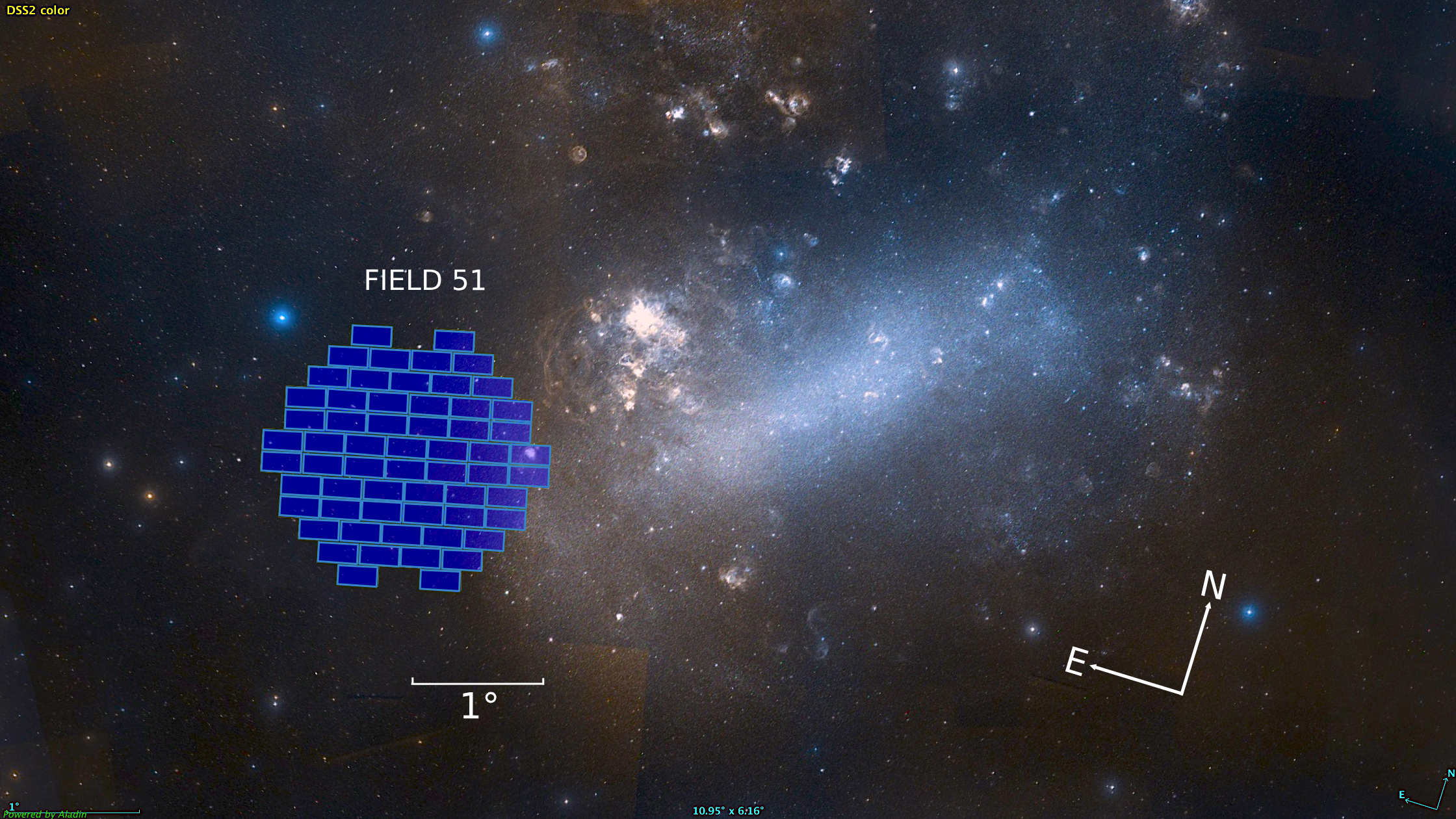}
\caption{LMC Field 51 shown on a $11^{\circ} \times 6^{\circ}$ area around the LMC. DECam FoV was generated using Aladin Sky with DSS2 colour images \citep{Aladin}.}\label{figFIELD51}
\end{figure}

AMPM (Asteroid-Mass Primordial black hole Microlensing) is an optical microlensing survey which aims to detect the lowest mass PBH microlenses in the Milky Way dark halo. In order to achieve such detections, a series of good signal-to-noise, rapid cadence observations is necessary to detect microlensing events with low finite-source amplifications and short Einstein crossing times ($t_{E} \sim$ minutes). We used the Dark Energy Camera (DECam) to image a single field in the LMC for the entire duration of the 5-night survey. The field, located at central coordinates $\alpha =$ 05:59:51.799, $\delta =$ -70:12:19.001 (J2000), coincides with Field 51 of the SMASH survey \citep{Nidever_2021}, which provided deep colour imaging in \textit{u}, \textit{g}, \textit{r}, \textit{i} and \textit{z} for the Large Magellanic Clouds. Positioned 2\textdegree45$'$ from the centre of the LMC, Field 51 offers a lower stellar density compared to the denser regions of the LMC central bar, see Figure \ref{figFIELD51} for the field's location. We intentionally selected a peripheral field for the survey to minimize photometric blending from crowded stellar regions, ensuring a high number of distinct stars for our microlensing search. AMPM images the LMC in a single, broad-band optical VR filter to avoid interruptions in the science observations caused by filter exchange overheads. The VR filter has a central wavelength of 626 nm and a full-width at half-maximum of 259 nm \citep{Decam_filter}, and allows for the deepest field observations with a short survey cadence when compared to the standard DECam filters. As the focus of AMPM is the asteroid-mass range of PBHs, a rapid imaging sequence is needed to be sensitive to fast transient signals. We use a 20-second exposure, which produces a cadence of 50 seconds when combined with the average read-out and overhead time of DECam.

\subsection{Photometry}
A variant of the  \texttt{DoPHOT} \citep{Dophot_1993} program was used to process exposures from the AMPM observations. The stellar PSF profile utilised for the photometry was a truncated power series, with the profile shape determined using a selection of bright stars. All fitted magnitudes on all images are aperture corrected using the procedures and rationale described in \citet{Saha_2010}. The correction, as implemented, adjusts for any slow variation of the PSF across the image area. The aperture-corrected magnitudes for the ensemble of objects from each image are then brought to a common zeropoint across the night, producing instrumental magnitudes on the same scale for all epochs. The instrumental photometry and other object attributes ascertained by \texttt{DoPHOT} were listed using a unique integer object identifier across all images on all nights. 

Finally, each night contains many millions of stellar light curves in the VR band, and we supplement the VR observations with colour information from SMASH DR2 to enhance our data. As AMPM is focused on fast microlensing events, we consider each of the five nights in the survey as an individual dataset and do not search across the combined nights. 

\section{AMPM SIMULATION GENERATION}
\label{sec:efficiency}
A key parameter of a microlensing survey is the efficiency ($\epsilon$), or the capability of any program to find candidate microlenses given the data, photometric conditions, and detection pipeline. Quantifying the efficiency is crucial for computing the expected event rate of the survey, which in turn is essential for constructing the PBH-DM limits. The raw efficiency is evaluated using the suite of simulated microlensing events and is defined as,
\begin{equation}
    \epsilon = \frac{N_{detect}}{N_{inject}}
\label{eq:effraw}
\end{equation}
Where $N_{detect}$ is the number of simulated events successfully detected through the AMPM pipeline, and $N_{inject}$ is the number of simulated microlensing events injected into the data. 

To evaluate the AMPM efficiency, we inject microlensing events into the observed light curves and compare the number of detected events to the number injected to generate an efficiency function as a function of PBH mass. 
The simulation suite considers the PBH distribution belonging to the Milky Way (MW) dark halo only, and generates events following the FS$-$PL microlensing via a direct computation microlensing model \citep{Lee_2009}. We do not model limb darkening in the FS$-$PL run as the $\sim 1 \%$ difference between the constant finite source and limb-darkened models is not measurable within the typical $\sim 1-5\%$ magnitude errors in the light curves. We use the simulated data to inform our selection of optimal detection thresholds for PBH microlenses.

\begin{figure}
\centering
\includegraphics[width = \columnwidth]{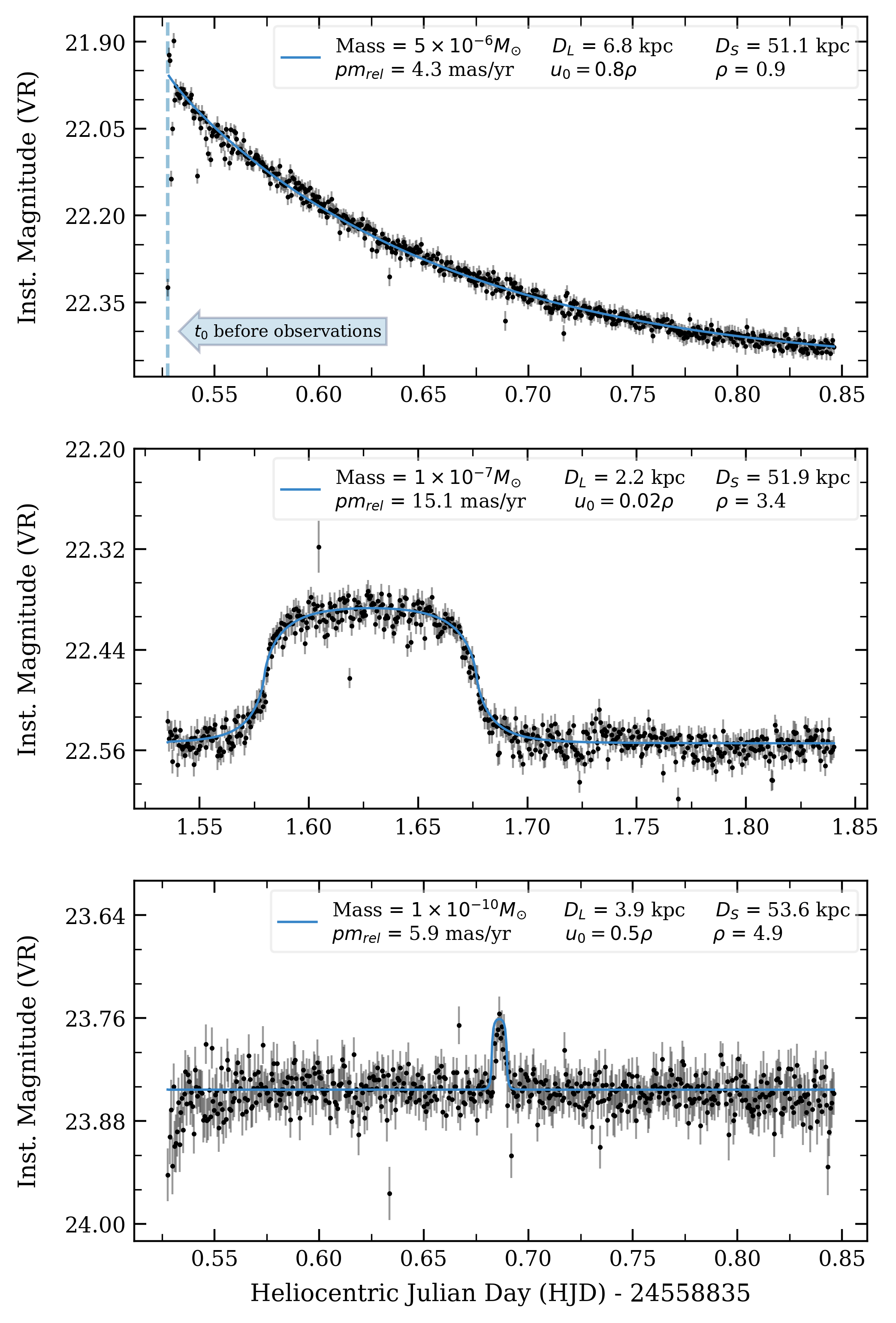}
\caption{Three examples of injected microlensing events for the FS$-$PL MW dark matter simulation. The variety of microlensing parameter combinations produces a wide range of signals; from the tail end of longer duration events (top panel), strong finite source damping events (middle panel), to low magnification, very fast events (bottom panel). The variety of microlensing events motivates the need for a flexible detection pipeline for the AMPM survey.}\label{figinjections}
\end{figure}

Microlensing events in the simulations are generated using \texttt{MulensModel} \citep{Mulensmodel}. For each night of the AMPM survey, we generate finite-source event iterations over a mass range of $10^{-13} - 10^{-3} M_{\odot}$, where every half decade in mass has 50,000 injected events. Each night contains 1.05 million simulated events, spanning the range of microlensing parameters based on the Navarro-Frenk-White (NFW) dark matter halo profile (as described in the next Section). The distances to the source star and lens ($D_{S}$, $D_{L}$) and transverse velocities ($V_{S}$, $V_{L}$) are drawn randomly from the respective MW and LMC distributions. The impact parameter ($u_0$) is randomly selected from $u_{0} \in [0.01, 1.1\rho]$ for the FS$-$PL case, where $\rho$ is calculated for each simulated FS-PL microlens. The peak time of the event ($t_0$) is randomly selected from an extended time range ($\Delta t$) around each light curve, defined by $\Delta t = t_{E} \times 0.5$ where $t_{E}$ is derived from the assigned mass, distances, and velocities of each event. We include the extended time buffer to encompass partially observed events in the AMPM efficiency. Figure \ref{figinjections} shows three examples of injected microlensing events with a range of masses and trajectory parameters.

\subsection{Milky Way Dark Halo}
We employ an NFW profile for the MW dark halo model \citep{Navarro_1997, Calcino_2018}
\begin{equation}
   \begin{aligned}
    \varrho_{DM}(r) = \varrho_{\odot}\frac{\frac{R_{\odot}}{R_{c}}(1 + \frac{R_{\odot}}{R_{c}})^{2}}{\frac{r}{R_{c}}(1 + \frac{r}{R_{c}})^2} \, .
\label{eq:7}
\end{aligned} 
\end{equation}
\newline\newline
The NFW parameters are taken from Table 2 of \citep{Calcino_2018} from a Markov Chain Monte Carlo (MCMC) analysis of galactic models fit to current Milky Way Rotation Curve data. The solar DM density is $\varrho_{\odot}=0.01058\,M_{\odot}\,\mbox{pc}^{-3}$, and scale radius $R_{c}=13.5\,\mbox{kpc}$. The isotropic velocity distribution for the Milky Way dark matter halo has a dispersion of $\sigma = 120 \,\mbox{km/s}$ \citep{Blaineau_2020, Battaglia_2005}, centred about a mean velocity of zero.

The LMC stellar distribution is modelled as a double exponential disk with a coordinate system $(x', y', z', R' = \sqrt{x'^2 + y'^2})$ that reprojects the origin to the centre of LMC Field 51, aligned with the galaxy's inclination, $i = 25.86^{\circ}$, and position angle $\theta = 149.23^{\circ}$ \citep{Sajadian_2021, Gyuk_2000}. The origin of the LMC as $(\alpha_{0}, \delta_{0}) = (82.25^{\circ}, -69.50^{\circ})$, and the central coordinate of Field 51 is $(\alpha, \delta) = (89.97^{\circ}, -70.21^{\circ})$,
\begin{equation}
\begin{aligned}
\varrho_{LMC}(R', z') =\frac{M_{d}}{4 \pi z_{d}R_{d}^{2}}\,\mbox{exp} \left(\frac{-R'}{R_{d}}\right)\,\mbox{exp}\left(\frac{-|z'|}{z_d}\right)\, ,
\label{eq:8}
\end{aligned}
\end{equation}
where the scale lengths are $R_{d} = 1.8  \mbox{kpc}$ and $z_{d} = 0.3 \mbox{kpc}$ 
and the central mass density $M_{d} = 2.55 \times 10^{9} \, M_{\odot}$. We take the distance to the LMC as 50 \mbox{kpc} \citep{Pietrzyski_2019}. The coordinate system transformation is described in detail in \citet{Sajadian_2021} and \citet{Weinberg_2001}. 
\newline\newline
We do not include a bar component in the LMC stellar density, as LMC Field 51 is sufficiently away from the central bar, and the disk stars dominate the density.  The existence of an LMC halo population has been discussed from measurements of heightened RR Lyrae velocity dispersions of $53 \pm 10$ \mbox{km/s} \citep{Alves_2004}, compared to the disc dispersion of $20.2 \pm 0.5$ \mbox{km/s} \citep{Marel_2002}. Stellar overdensity mapping from SMASH photometry also indicates the presence of a low-density halo contribution at large radial distances that comprises $\sim0.4\%$ of the total stellar mass \citep{Nidever_2019}. As Field 51 is $2.75^{\circ}$ degrees from the LMC centre, which exists within the disk density profile range within $15^{\circ}$ \citep{Nidever_2019}, and the disk density significantly dominates the stellar content, we do not model an LMC stellar halo. The stellar velocity distribution is drawn from \citep{Kallivayalil_2013} with a galactocentric tangential velocity of $<v> = 314\,\mbox{km/s}$, and a dispersion of $\sigma = 24 \,\mbox{km/s}$.

\subsection{Determining Stellar Radii for FS--PL Injection}
\label{ssec:Rstars}
An estimate of the source radius ($R_{S}$) is necessary to build the amplification curve of a FS$-$PL microlensing event. Since AMPM did not capture multi-filter images, we estimate the source radius using archival data and by assuming all stars reside within the LMC. We follow the method described by \cite{Smyth_2020} and utilise \texttt{MIST} synthetic photometry to model LMC sources and their radii \citep{dotter_MIST, choi_MIST, Paxton_MISTA, Paxon_MISTB, Paxon_MISTC}. 

First, we retrieve all Field 51 stars from the SMASH DR2 catalogue and match them to the nearest AMPM stars to assign \textit{g} and \textit{r} magnitudes, with a median cross-match offset of 0.2 arcseconds. Next, we generate $10^{7}$ synthetic track samples using MIST with the typical LMC metallicities between $-$1.0 and 0.5 dex [Fe/H], $\log_{10}$ ages between 6.5 and 9, and a Chabrier mass distribution between $1-80 M_{\odot}$ \citep{Narloch_2022}. Synthetic photometry catalogues are produced from the track samples for DECam \textit{g}- and \textit{r}-band using the \texttt{isochrones} package. These catalogues include magnitude measurements and stellar properties, such as radius and effective temperature. The synthetic photometry is then matched to the augmented SMASH+AMPM catalogue by a nearest neighbour query on the \textit{g}-magnitude and \textit{g}-\textit{r} colour, with a median and standard deviation on the match separation of $0.002 \pm 0.133$ mag. Finally, each AMPM star is assigned the synthetic stellar radius of its closest match from the MIST photometry. The synthetic stellar radius is used during the efficiency simulation to generate FS-PL microlensing events.
\begin{figure}
\centering
\includegraphics[trim={0 0 0 0},clip,width=1.\columnwidth]{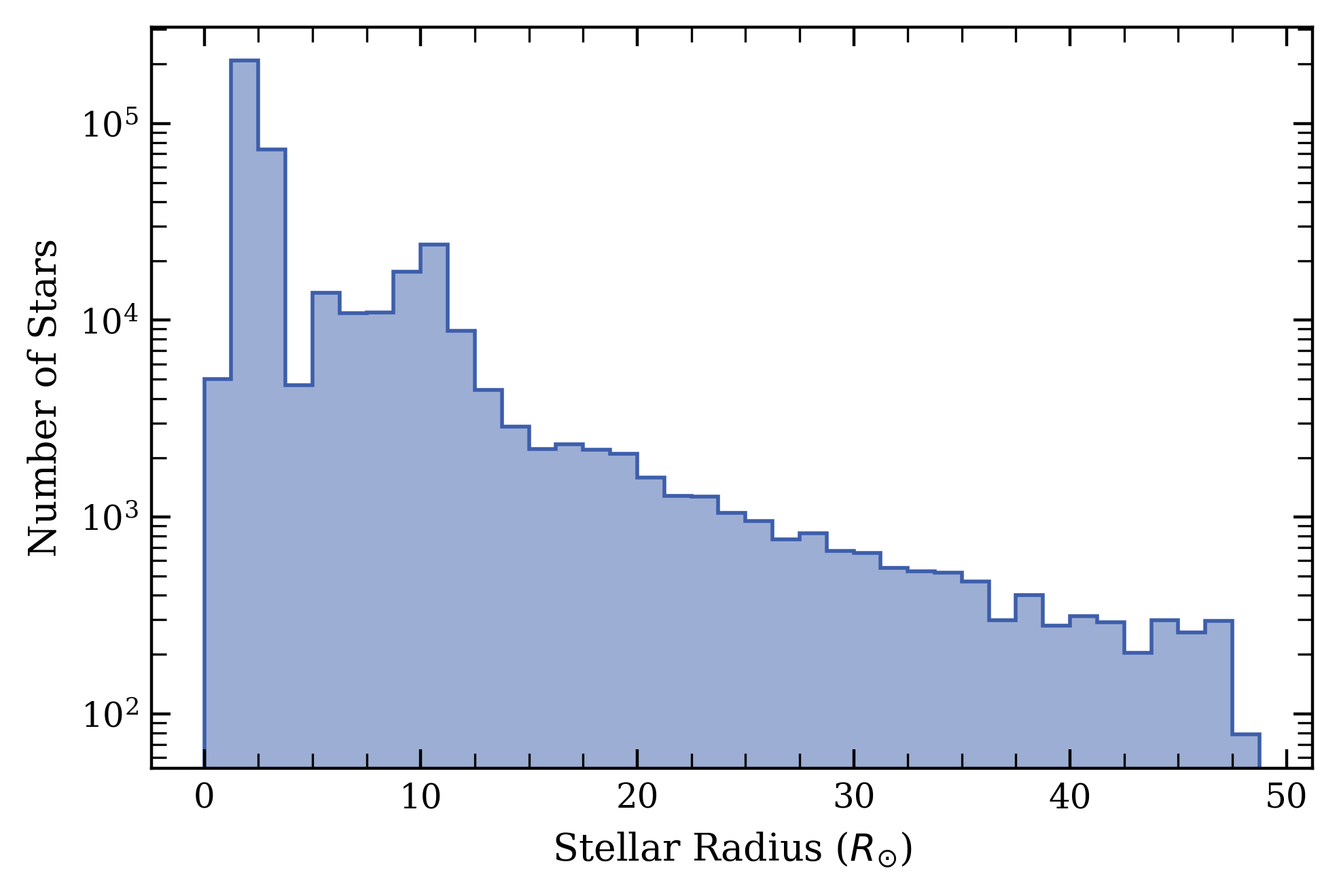}
\caption{The distribution of stellar radii from the synthetic MIST photometry compared to the SMASH \textit{g} and \textit{r} magnitudes for Field 51. A broad distribution of star sizes in Field 51 exists. Most stars are main sequence analogues with $R_{\mbox{star}} \gtrsim 1 R_{\odot}$, but the field also contains a significant number of $10 R_{\odot}$ stars and giant stars larger than $40 R_{\odot}$. Since many LMC stars are larger than $1 R_{\odot}$, the microlensing events in AMPM will have a pronounced finite source dampening effect.}\label{figureREfstars}
\end{figure}

The average distribution of Field 51 radii across the 5 nights of the survey is shown in Figure \ref{figureREfstars}. While many main sequence stars with $\sim 1 R_{\odot}$ exist in the AMPM catalogues, giant stars of order $O(10) R_{\odot}$ are also present in large numbers. The high proportion of giant stars in the AMPM catalogues leads to a noticeable reduction in peak amplification for events with high $\rho$ values, resulting in lower efficiency for these events. This efficiency shift was similarly noted in the OGLE-IV LMC PBH survey, where an increase in $\log_{10} \rho$ from $-3$ to $0.8$ shifts the 1\% efficiency bound from a sensitivity to $t_E \sim 0.01$ to $t_E \sim 0.1$ days \citep{Mroz_OGLEIV2024}. Consequently, finite source effects are the most important contribution to the efficiency of an asteroid-mass microlensing survey. 

\subsection{An Absolute Mass Limit from Wave Optics}
The treatment of microlensing changes when considering the light from a source star as a monochromatic wave \citep{Nakamura_1999}. Rather than the microlens deflecting light rays, now the lens bends and alters the direction of the flux wavefront emitted from the source, and causes diffraction patterns of the lensed images in time. When the physical size of the lens is comparable to the wavelength of the source light, the diffraction effect in microlensing must be folded into the amplification calculation. As the focus on constraining PBH-DM abundance moves towards the lower mass regimes, the classical treatment of microlensing --- where the lens system yields coherent images --- is no longer valid. The event amplification is further damped by wave optics, with an additional $A_{max}$ effect that is distinct from the finite source effect \citep{Sugiyama_2020} given by,
\begin{equation}
    A^{wo}_{max} = \frac{\pi \,\mathrm{w}}{1 - e^{-\pi\,\mathrm{w}}} \, ,
\end{equation}
which depends on the angular frequency of light ($\mathrm{w}$) defined as \citep{Matsunaga_2006}.
\begin{equation}
       \mathrm{w} = \frac{\omega\,D_{S}\,D_{L}\,\theta_{E}^2}{c\,(D_{S} - D_{L})} \equiv \frac{8\,\pi\,G\,M}{c^{2}\,\lambda} \, .
\end{equation}
For a microlensing system with a set geometry and lens mass, the observed wavelength of light alters the maximum detectable magnification of the source wavefront. We take the $A^{wo}_{max}$ detectable threshold to be 1.085, and find the corresponding minimum visible lens mass for the AMPM survey using the central wavelength of VR filter $\lambda$ = 626 nm 
as $9 \times 10^{-13} M_{\odot}$. 


\section{The AMPM Microlensing Detection Pipeline} \label{sec:detect}
In this section, we describe the process of detecting microlensing signals in the AMPM dataset by using an optimised set of statistics tailored to discover trended light curves. 

\subsection{Quality Control} \label{ssec:QualCont}
An initial stage of quality control is applied to the data to identify the light curves with the most complete and reliable measurements. The total completeness of a light curve doesn't necessarily damage the detection efficiency; light curves that lose every alternate measurement effectively mimic observations with a doubled cadence. Rather, it is clusters of missing data points that damage the detectability of an event. So, we employ two universal criteria that remove poor quality light curves from each catalogue.

\begin{figure}
\centering
\includegraphics[width = \columnwidth]{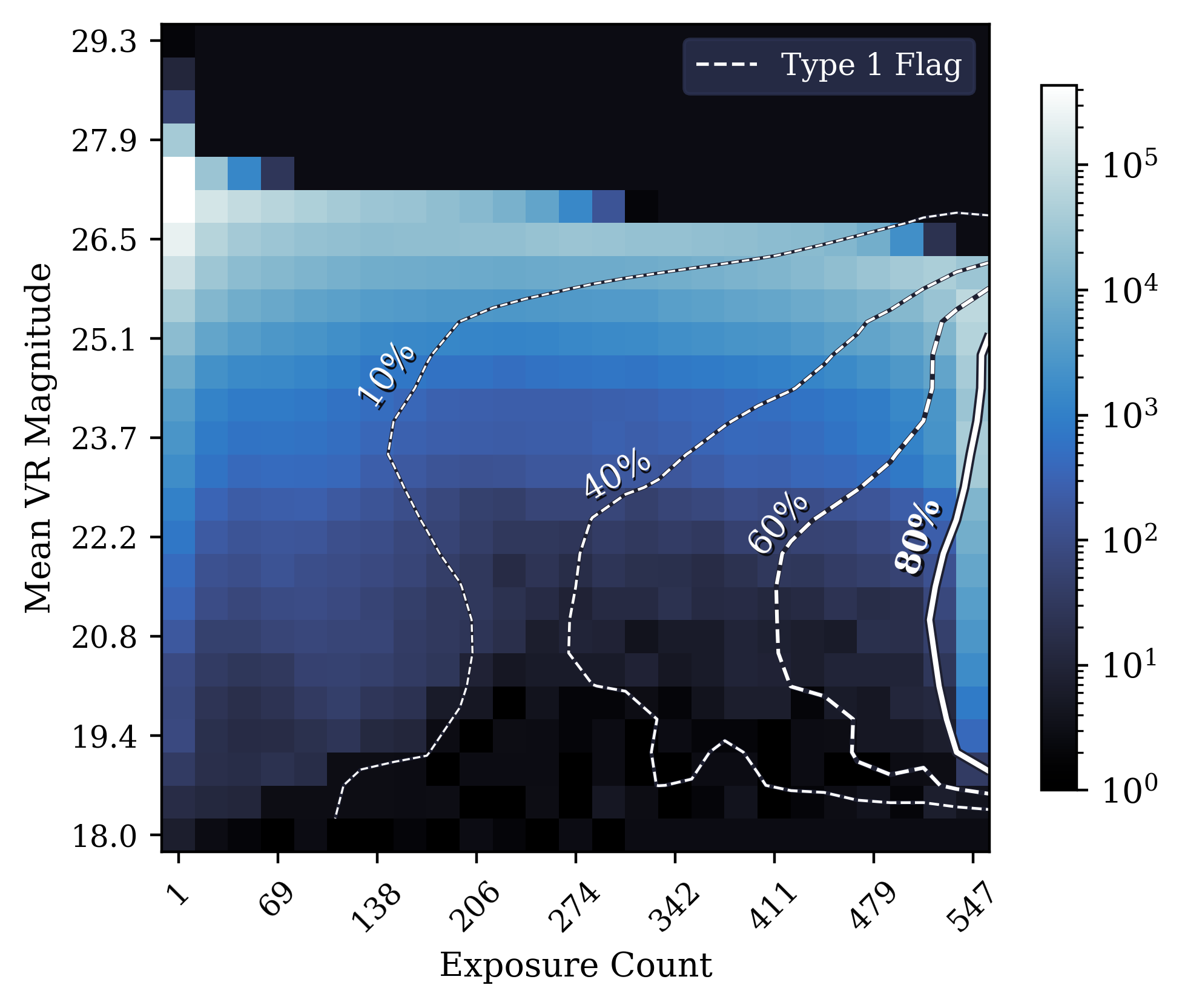} 
\caption{2D Distribution of sources in the parameter space of mean instrumental magnitude (VR-band) of the AMPM survey and the number of images per light curve in December $15^{\rm th}$. Contours show the percentage distribution of DoPHOT \textit{Type 1} values per light curve. Higher percentages of \textit{Type 1} denote well-fitted stellar objects. Clearly, low \textit{Type 1} < 10\% trace the region of fainter and poorer quality light curves. The 80\% \textit{Type 1} contour corresponds to the AMPM quality control query.}\label{figTYPE}
\end{figure}

The first criterion uses a DoPHOT flag generated during the photometry. 
DoPHOT’s \textit{Type} parameter is a numerical value that indicates the confidence that a star has been accurately fitted by the PSF model, and is neither an extended object nor contamination by image artifacts. We retain only the light curves classified as well-fitted stars (\textit{Type 1}) for over 80\% of the night. Figure \ref{figTYPE} illustrates the source distribution for December $15^{\rm th}$ against the mean VR magnitude and the total sum of the data points in the light curves (with a maximum of 570 exposures). The \textit{Type} parameter effectively identifies complete light curves without imposing a hard limit on the minimum number of data points or limiting magnitude. 

The second criterion removes the light curves that pass the \textit{Type} criterion but are located close to the edges of the CCD chips across the DECam plane. Maintaining a single target field necessitates repointing DECam throughout the night, which leads to slight variations in the chip orientation relative to Field 51. Sources within 40 pixels from the CCD's edge contain significant portions of missing measurements as they fall off the detector plane and into the intra-CCD spacing that spans of order 100 pixels. This Gap Criterion removes a further five percent of the remaining sources. 

We then search the remaining 2 million light curves for events as short as a few minutes to several hours in duration. In the case of AMPM, as no real-time detection is required, the most straightforward method is the staged, cut-based pipeline. For the AMPM survey, the finite source effect dampens the peak amplification of microlensing events towards the LMC, and many events will not rise significantly above the light curve's baseline magnitude and standard deviation. Using a classic statistical threshold of an arbitrarily-selected number of 3$\sigma$ consecutive data points to query events in the catalogue will not be conducive to finding fast FS$-$PL microlenses.
\subsection{Von Neumann Statistic}
The Von Neumann ratio ($\eta$) represents the mean squared successive differences between data points, normalized by the variance, and is used to test whether the data are stationary \citep{VN1_1941, VN2_1941}. The use of the $\eta$ statistic has risen in microlensing as it is efficient to compute for large influxes of data in both statistical and machine learning analyses \citep{PriceWhelan_2014, Godines_2019, Husseiniova_2021, Kim_2014}. The Von Neumann ratio is defined as
\begin{equation}
    \eta = \frac{n}{(n - 1)}\frac{\sum\limits^{n-1}_{i = n}(m_{i+1} -  m_{i})^{2}}{\sum\limits^{n}_{i = n}(m_{i} -  \overline{m})^{2}}
\end{equation}
where $m_{i}$ are the individual magnitude measurements and $\overline{m}$ is the mean magnitude of the light curve. For uniformly distributed homogeneous data, $\eta$ = 2 with $\eta$ tending towards zero as trends in the data become stronger \citep{Press_1969}.
\subsection{Seeing-related Flux Blending \textit{(B)}}
In crowded stellar fields, poor atmospheric observing conditions can artificially expand the stellar PSF to encompass neighbouring stars, confounding the ability to detect the true variability of the star. The PSF overlap induces atmospheric blending and results in an averaged fluctuation pattern that is observable in many light curves. However, the strength and manifestation of the pattern may vary with the brightness and crowding of the stars. Unfortunately, these fluctuation patterns occur over the same timescales as the microlensing events of interest to the AMPM survey. The average full-width half-maximum duration of the seeing patterns is approximately eight minutes, which is comparable to the mean $t_{E}$ duration for $10^{-9} - 10^{-8} M_{\odot}$ PBHs towards the LMC (see Section \ref{sec:efficiency}).

\begin{figure}
\centering
\includegraphics[trim={0 0 0 0},clip,width=1.\columnwidth]{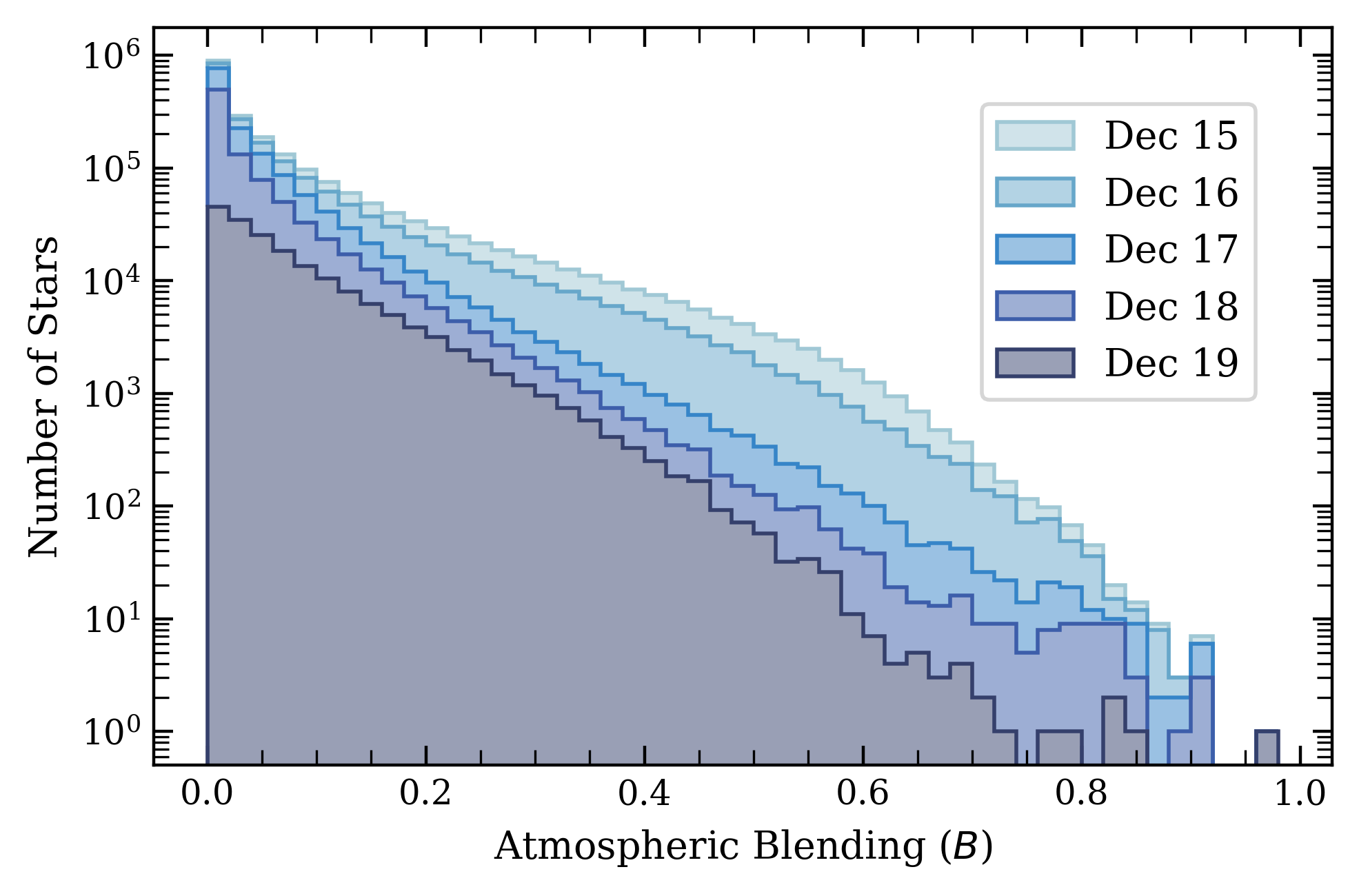}
\caption{The distributions of the \textit{B} parameter measuring the correlation of light curve fluctuations to atmospheric seeing across each night of the AMPM survey. Each histogram represents the distribution of catalogue stars in that night. A $B$ value approaching zero indicates there is no evidence for atmospheric blending. High values of \textit{B} correspond to greater temporal seeing patterns in each light curve. December $15^{\rm th}$ and $16^{\rm th}$ contain more seeing contamination as indicated by the increase of sources at $B > 0.3$}\label{figBhist}
\end{figure}

\citet{Irwin_2007} describe a method for quantifying the amount of seeing-related variability in light curves through a comparison between the $\chi^{2}$ result from a constant, non-varying magnitude model to that of variability correlated with the change in the PSF size. The result from the PSF modelling is a parameter referred to as \textit{B}.
\begin{equation}
    B = \frac{\chi^{2}_{\mbox{const}} - \chi^{2}_{\mbox{fit}}}{\chi^{2}_{\mbox{const}}}
\end{equation}
The $\chi^{2}_{\mbox{fit}}$ parameter is measured from the best-fit quadratic function in the $\Delta m = (\overline{m} - m_{i})$ and PSF FWHM space, and $\chi^{2}_{\mbox{const}}$ is measured from a $y = \overline{m}$ model. The \textit{B} parameter is like a Bayesian evidence parameter that measures the statistical significance of correlations with the seeing pattern in the night. As the resulting value of the $\chi^{2}_{\mbox{fit}}$ parameter reduces, \textit{B} tends towards a maximum of 1, and the variability in the light curve is better explained by the variation in the PSF FWHM. We evaluate the average PSF FWHM in each image by selecting 30 bright, isolated and symmetric stars across three CCDs\footnote{CCDs S29, S7 and N22, which are respectively among the northernmost, easternmost and central chips, in order to sample stars across the breadth of the detector plane.}. Each star is fitted with a Gaussian PSF model, and the final PSF FWHM for the image is measured as the median of the subset of stars. The distributions of the \textit{B} parameters for all five nights of the survey are shown in Figure \ref{figBhist}. The increase of stars with $B > 0.3$ for December $15^{\rm th}$ and $16^{\rm th}$ shows that these nights suffer worse seeing, and hence PSF blending, for more stars than the final three nights. 

\subsection{Smoothed Peak Detection \textit{(P, Max, Min)}} \label{sec:smooth}
Although the Von Neumann ratio is an efficient statistic in quantifying variability, it doesn't provide the exact location or duration of the variability pattern in a light curve. Suppose the light curve has strong atmospheric contamination. In that case, the \textit{B} parameter only provides implicit temporal information about the trend in a light curve (i.e. the trend follows the PSF fluctuations over the night). Therefore, we construct a function that combines the optimal smoothing of each light curve with a change-point function to locate peaks in the light curve. 

We use the algorithm presented in \cite{Garcia_2010}, which automatically computes the optimal smoothing length of data via minimisation of the Generalised Cross-Validation score (GCV), and then smooths the data using discrete cosine transforms in the penalised least squares method. The optimal smoothing algorithm (hereafter referred to as G10) is most often used in transient detection for smoothing of data to remove and detrend variability \citep[see eg.][and references therein]{Li_2019, Thompson_2018}. We use the G10 method to smooth the noise fluctuations in each light curve in preparation for finding peaks. The smoothing length, $S$, is defined as
\begin{align}
&S = \mathrm{argmin(GCV)} \, ,\\
&\mathrm{GCV}(S) = n\frac{\sum_{i = 1}^{n}(\hat{m_{i}} - m_{i})^2}{(n - \sum_{i = 1}^{n}(1 + S\lambda_{i}^{2})^{-1})^{2}} \, , \\
&\lambda_{i} = -2 + 2\cos\left(\frac{(i-1)\pi}{n}\right) \, ,   
\end{align}
where $\hat{m}$ is the smoothed estimation of the light curve $m$, and $n$ the number of data points in the light curve. The GCV is minimised using the logarithm of $S$. Local maxima (\textit{Max}) and minima (\textit{Min}) are found with an extrema analysis on the G10 smoothed light curve. The maxima and minima locations are used to compare the frequency of peaks in a light curve and the duration between repeated peaks.

\subsection{Periodicity Time Comparison \textit{(T)}} 
Microlensing by isolated PBHs is non-repeating. We reject a microlensing candidate if another comparable signal is detected in its light curve.Some classes of variability can be mistaken for fast microlensing events (such as the optical signal from flaring stars), but periodic variable stars can be easily detected from the repeated variability pattern. We define a statistic, \textit{T}, which is used to classify the strength of the predominant period of variability in a light curve. The time between the peaks (\textit{Max}) previously detected from the smoothed peak detection algorithm is calculated as a naive variability period ($\mbox{T}_{PK}$) by computing the median difference in the Heliocentric Julian day of the exposures corresponding to peaks. A second period ($\mbox{T}_{LS}$) is evaluated as the inverse of the maximum power frequency from the Lomb Scargle periodgram \citep{Lomb_1976, Scargle_1983} as implemented by \texttt{Astropy}. 
The normalised difference between the two periods is denoted as \textit{T} and given by
\begin{equation}
T = \frac{|\mbox{T}_{LS} - \mbox{T}_{PK}|}{\mbox{T}_{LS}}.
\end{equation}
The \textit{T} parameter is a straightforward way of representing the concordance between a frequency-space and temporal-space period analysis. For large values of \textit{T}, the Lomb-Scargle analysis has identified long underlying variability often of several days, while the peak detection will usually detect small noise-like fluctuations. As \textit{T} approaches zero, the variability expressed in the light curve becomes stronger and quicker with periods of a few hours in duration.


\subsection{Trend Detection with Statistics} \label{ssec:Trend}
\begin{figure}
\centering
\includegraphics[trim={0 0 0 0},clip,width=1.\columnwidth]{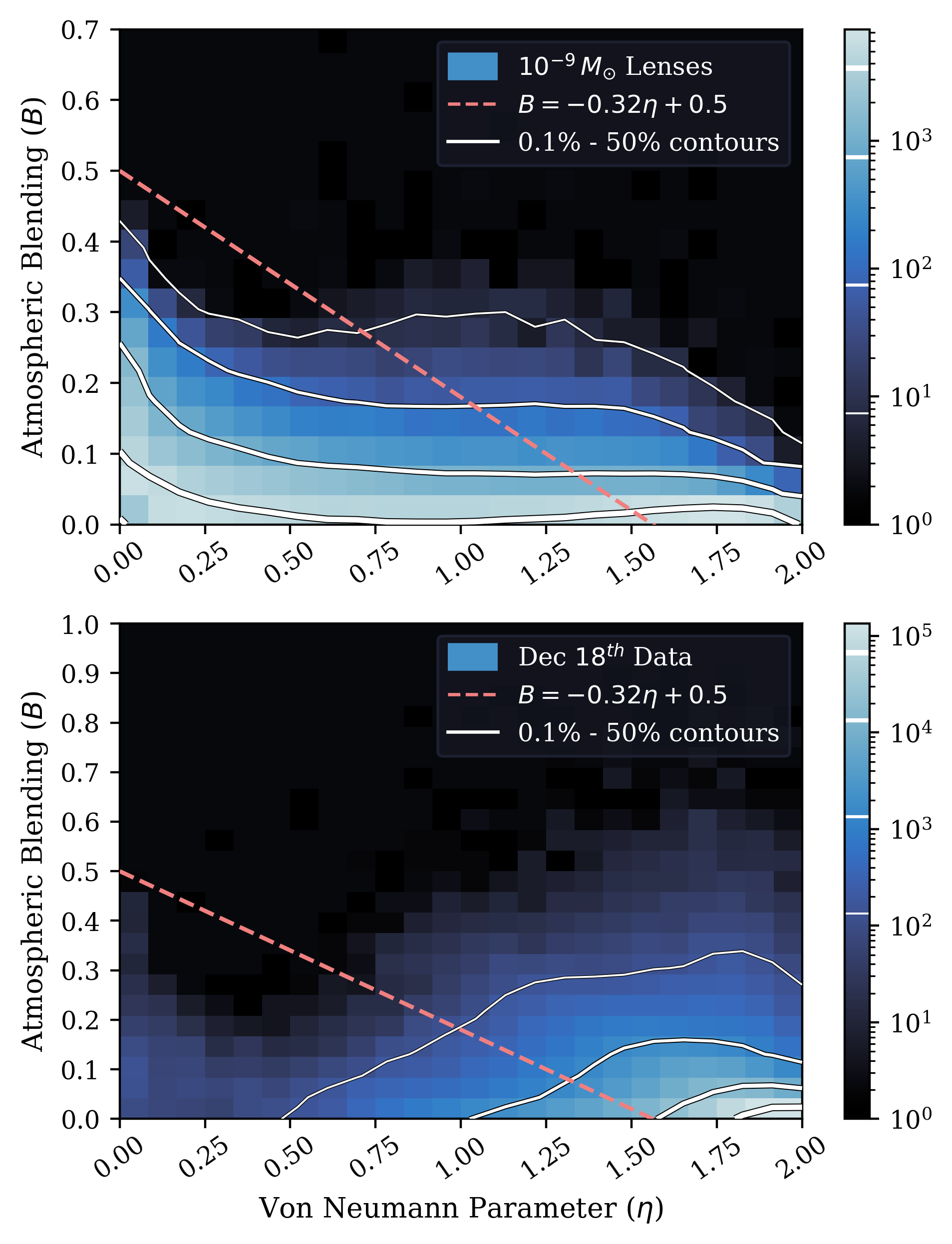}
\caption{The AMPM trend detection threshold from Equation \ref{eq:detectthresh} for the Seeing-related Flux Blending parameter, $B$. The threshold is shown as the pink dashed line superimposed on the set of simulated microlensing events of $10^{-9} M_{\odot}$ mass in the top panel, and over the set of December $18^{\rm th}$ data. The concentration of simulated microlenses at low values of blending and Von Neumann parameter traces the relevant interesting regions for the parameter space of December $18^{\rm th}$.}\label{threshB}
\end{figure}

\begin{figure}
\centering
\includegraphics[trim={0 0 0 0},clip,width=1.\columnwidth]{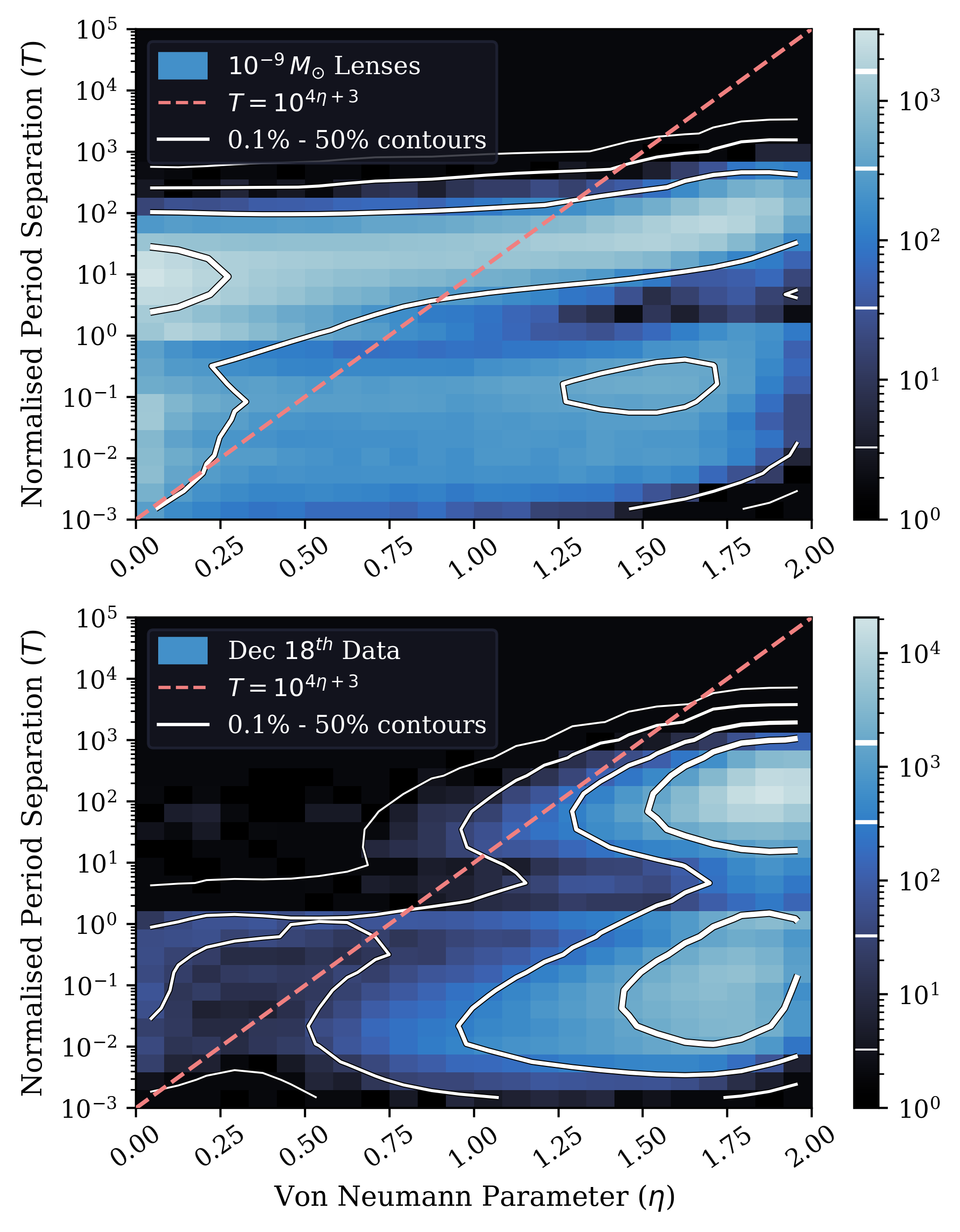}
\caption{Similar AMPM trend detection threshold from Equation \ref{eq:detectthresh} for the Periodicity Time Comparison, $T$. Again, the threshold is shown as the pink dashed line superimposed on the set of simulated microlensing events of $10^{-9} M_{\odot}$ mass in the top panel, and over the set of December $18^{\rm th}$ data.}\label{threshT}
\end{figure}

As each microlensing statistic is tailored to distinguish different variability patterns, we construct a microlensing candidate detection threshold in the joint $\eta$, \textit{B} and \textit{T} parameter spaces. Using all parameters together improves the identification of trended light curves. For instance, stars with high \textit{B} but very low $\eta$ values likely show genuine trends and should be retained. The threshold for detecting microlensing signals is calibrated and validated through synthetic microlensing event injection, as described in Section \ref{sec:efficiency}. The microlensing detection threshold is formally defined as,
\begin{equation}
    (\eta \leq 1) \; \And \; (B \leq B_{\eta}) \; \And \; (T \leq T_{\eta}) \, ,
\end{equation}
where
\begin{equation}
B_{\eta} =   -0.32\eta + 0.5, \qquad
T_{\eta} =   10^{(4\eta - 3)} \, .
\label{eq:detectthresh}
\end{equation}
Figure \ref{threshB} and Figure \ref{threshT} include the $B_{\eta}$ and $T_{\eta}$ boundaries on the 2D distribution of stars from December $18^{\rm th}$ (4th night) and simulated microlensing events injected into December $18^{\rm th}$ light curves. The boundaries are selected from the December $18^{\rm th}$ simulations to maximise our microlensing event detection efficiency. 
\subsection{Repeated Variability Rejection} \label{ssec:RepVar}
Light curves that pass the threshold condition are then passed onto a second variability analysis. Successful candidate events are checked against the Gaia DR3 variability catalogue for sources within Field 51 \citep{GaiaVarInfo}. If a variable star within 0.5 arcseconds is known, the candidate is removed from consideration. The light curves for each candidate are selected from the quality catalogues and are evaluated for variability using the $\eta$ statistic. If the median $\eta$ for the additional nights is less than one, the light curve is removed from the candidate list. This second stage is aimed at removing variable stars with archival long-duration variability and predictable, rapid variability within the AMPM survey.

\subsection{Microlensing Fitting Analysis} \label{ssec:Muchi}
\begin{figure*}
\centering
\includegraphics[trim={0 0 0 0},clip,scale = 0.8]{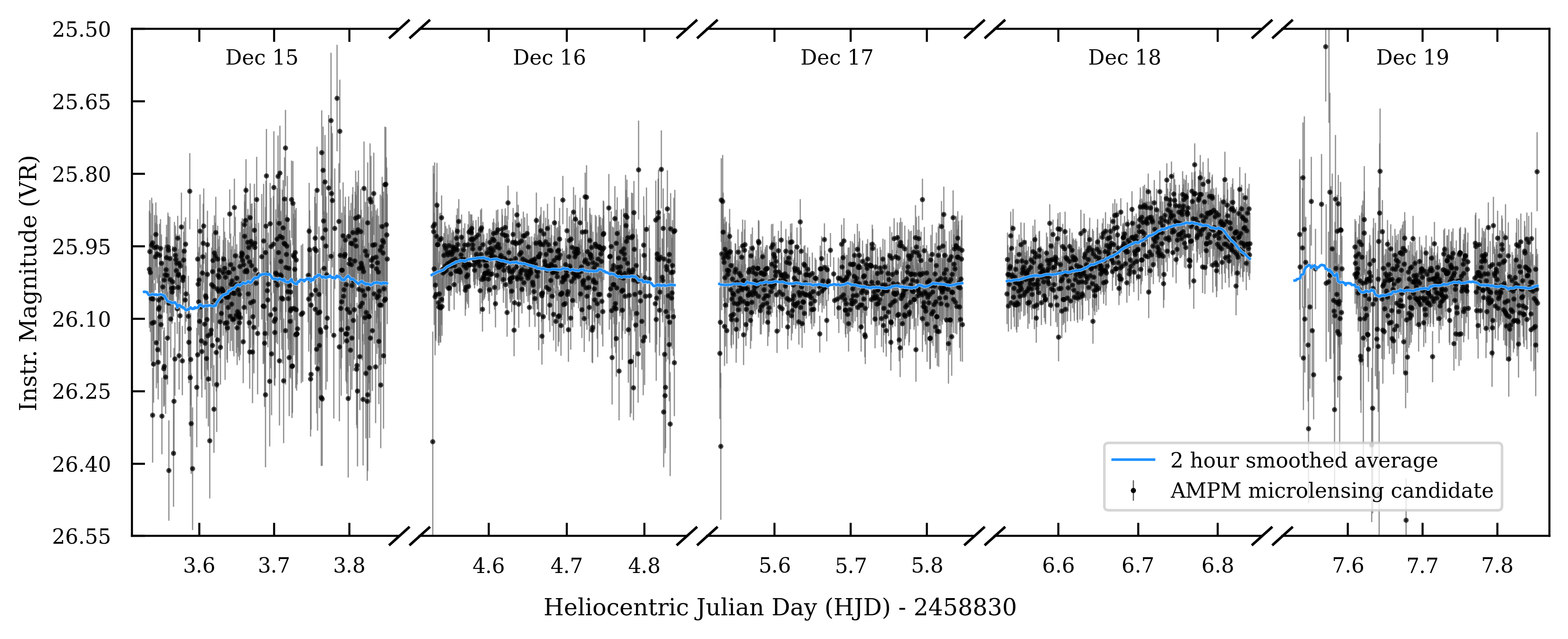}
\caption{The detection light curve for the PBH candidate from December $15^{\rm th} - 19^{\rm th}$. Inferior observing conditions on December $15^{\rm th}$ produce strong scatter in stellar magnitudes that significantly affect all stars in the field. Similar localised sections of atmospheric scatter are evident in the light curve towards the end of December $16^{\rm th}$ and the beginning of the night on December $19^{\rm th}$. The smoothed signal plotted in blue is evaluated using a box kernel with a 2-hour window, or equivalently a window of 120 datapoints. }\label{figphoebe}
\end{figure*}

We evaluate microlensing candidates using a fast PS$-$PL fitting routine. While the most realistic and accurate microlensing treatment would be with FS$-$PL modelling, the analytical evaluation of finite-source effects greatly increases the computation time for the fitting routine. The PS$-$PL method prioritises quick processing while still providing adequate model fitting to the data. Firstly, initial estimates of the PS$-$PL microlensing parameters $t'_{0}, u'_{0}, t'_{E}$ are evaluated from the smoothed light curve, $m_{i}$, from the Smoothed Peak Detection in Section \ref{sec:smooth}. 
\begin{align*}
&t'_{0} = \mathrm{HJD}(m_{i} = \mathrm{min}(m_{i})) \, , \\
&u'_{0} = 10^{-0.4(\overline{m_{i}} - \mathrm{min}(m_{i}))} \, , \\
&t'_{E} =\frac{t_{FWHM}}{2} \, , \\
\end{align*}
where $t_{FWHM}$ is defined as the number of consecutive data points $1\sigma$ below the smoothed mean magnitude $\overline{m_{i}}$. Using \texttt{MulensModel}, a bounded minimiser evaluates the best PS$-$PL amplification curve defined from Equation \ref{eq:Aps}. The reduced $\chi^{2}$ values of the PS$-$PL model is returned as statistics for the AMPM pipeline, as well as the maximum peak magnitude ($dm$) and magnitude errors ($err_{dm}$) of the signal. We use three concurrent conditions to isolate potential microlensing candidates; a good reduced $\chi^{2}$ value, a peak that is not negligible above the magnitude noise, and a significant brightening with the best-fit magnitude and errors ($\sigma_{dm} = dm/err_{dm}$),
\begin{equation}
    (\chi^{2}_{red} < 1.5) \,\, \& \,\,(\chi^{2}_{red} > 0.5) \,\, \& \,\,(dm < -0.05) \,\, \& \,\,(\sigma_{dm} > 10).
\end{equation}
Finally, all remaining microlensing candidates are visually inspected. Most of the remaining candidates are fast flare stars, low amplitude and faint variable stars, and eclipsing systems with variability appearing as the tail end of extreme finite source microlensing. The reduction of candidates throughout the AMPM pipeline is detailed in Table \ref{tab:candidates}. We find a single microlensing candidate on the fourth night from the AMPM pipeline, with the full five nights of data shown in Figure \ref{figphoebe}. The candidate has a full-width half-maximum timescale of $t_{FWHM} = 3.57$ hours. We present, analyse, and extensively discuss the impact of the single PBH candidate from AMPM in the second paper of this series, AMPM II. -- A Lunar-Mass Primordial Black Hole Microlensing Candidate in the Milky Way Halo.

\renewcommand{\arraystretch}{1.5}
\begin{table*}
\small
\begin{tabular}{lrrrrrc}
\hline
\textbf{Stage} & \textbf{Dec 15} & \textbf{Dec 16} & \textbf{Dec 17} & \textbf{Dec 18} & \textbf{Dec 19} &\textbf{Median Remaining Sources ( \%)}\\
\hline
DoPHOT Catalogues & 3,289,506 & 3,895,884 & 4,075,224 & 4,092,824 & 3,824,668 & 100\\
Post Quality Control (sec. \ref{ssec:QualCont}) & 248,656 & 381,418 & 553,324 & 696,942 & 184,807 & 10.77\\
Post Trend Detection (sec. \ref{ssec:Trend}) & 6,873 & 15,132 & 9,417 & 9,136 & 2,378 & 0.22\\
Post Repeated Variability Rejection (sec. \ref{ssec:RepVar}) & 5,000 & 12,338 & 5,473 & 5,146 & 1,452 & 0.15\\
Post Microlensing Fitting Analysis (sec. \ref{ssec:Muchi}) & 143 & 5,559 & 3,545 & 3,080 & 185 & 0.0006\\
\hline
Microlensing Candidates  & 0 & 0 & 0 & 1 & 0\\
\hline
\end{tabular}
\caption{Source Counts at representative stages through the AMPM microlensing detection pipeline, from the raw input catalogues to the final number of microlensing candidates in each night of the survey.}
\label{tab:candidates}
\end{table*}


\section{Limits on PBH Dark Matter} \label{sec:limits}
\begin{figure}
\centering
\includegraphics[trim={0 0 0 0},clip,width=1.\columnwidth]{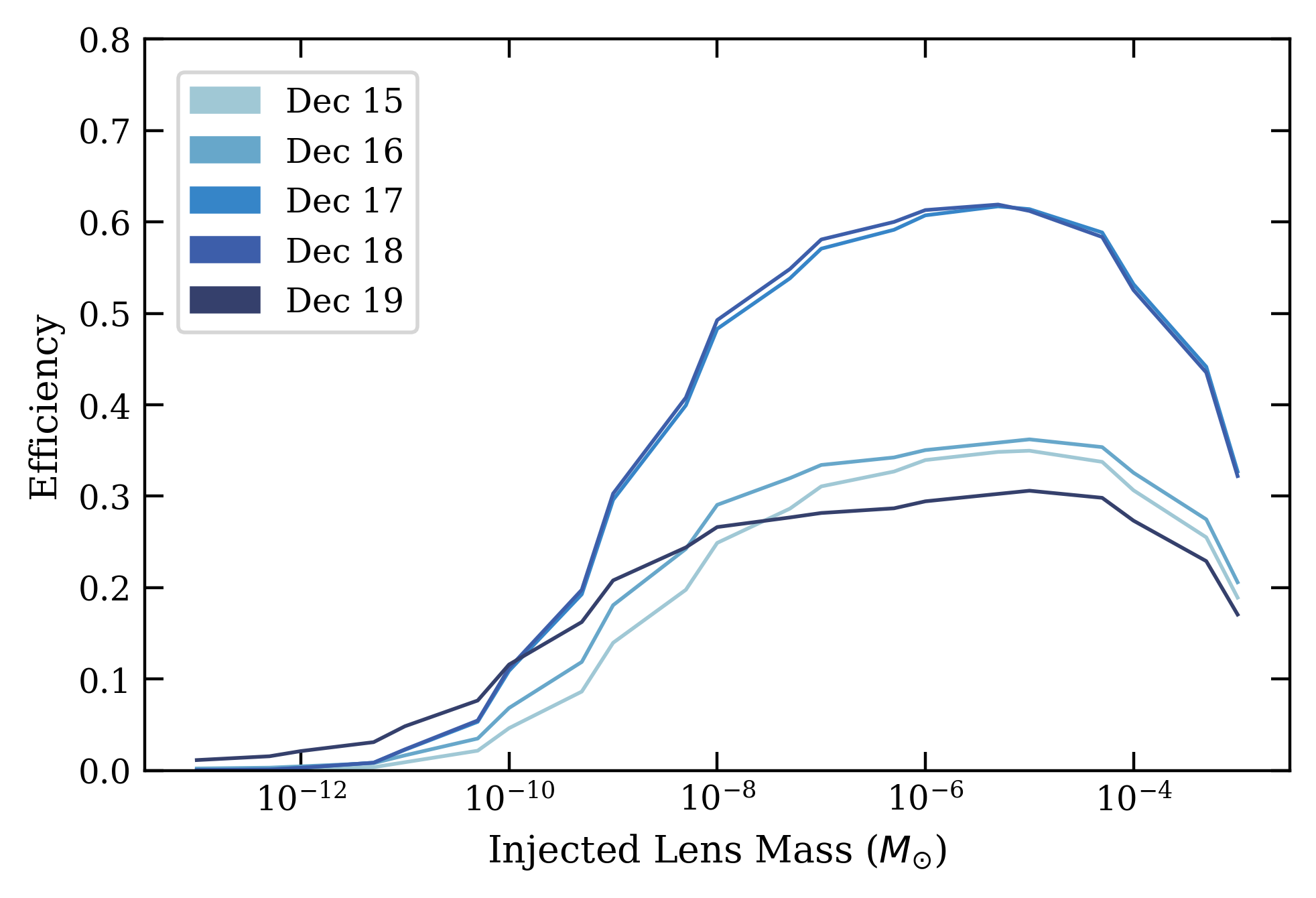}
\caption{The efficiency of the AMPM microlensing detection pipeline over the five nights of the survey under a realistic finite source, point lens (FS$-$PL) treatment for the LMC stellar distribution. The microlensing simulations are generated according to the description in Section \ref{sec:efficiency}.}\label{fignighteff}
\end{figure}

In order to place constraints on the fraction of the dark matter density in PBHs, we calculate the number of microlensing events expected for the Milky Way NFW halo with a PBH dark matter density fraction $f_{PBH} = 1$. The number of expected events $N_{exp}$ is defined as the product of the detection efficiency weighted rate of microlensing, $\Gamma$ in units of star$\cdot$days$^{-1}$, with the duration of the survey $T$ days, and the number of stars monitored $N_{S}$,
\begin{equation} \label{eq:nexp}
     N_{exp} = T \; N_{S} \; \epsilon(M)\;\Gamma(M)
\end{equation}
and the event rate is the ratio between the optical depth -- the probability of a single star being microlensed by an object of mass $M$ -- and the mean event timescale at that mass \citep{Calchinovati_2013, Tisserand_2007}.
\begin{equation}
    \Gamma(M) = \frac{\tau}{<t_{E}(M)>}
\end{equation}
The efficiency of the AMPM pipeline is calculated using Equation \ref{eq:effraw} over each suite of simulated events of the five nights of data. 
As shown in Figure \ref{fignighteff}, the maximum efficiency is for PBH masses of $10^{-5}\, M_{\odot}$. The efficiency is naturally lower for PBHs below $\sim 10^{-8}\, M_{\odot}$ as the $A_{max}$ amplification damping to the FS$-$PL event is stronger at lower masses. For the nights with the best observation conditions, December 17-18, the maximum efficiency is $\epsilon = 0.6$. The three remaining nights show clearly that poorer seeing conditions strongly affect the efficiency, with the maximum being around $\epsilon = 0.3$. 

We assume a monochromatic mass distribution of PBHs, such that $\Gamma(M)$ is evaluated for each PBH mass sampled in the efficiency simulation. For the FS$-$PL evaluation, we calculate the optical depth by including the $D_{max}$ condition in the integral bounds as in Equation \ref{eq:tau}. The $D_{max}$ condition acts to reduce the sightline over which the optical depth is integrated, and hence reduces the probability of seeing very low-mass microlenses. We incorporate the stellar distribution of Field 51 by considering the histogram of the stellar radii, in steps of $0.5 R_{\odot}$. For each bin, we use the centre value for the radius to evaluate the $D_{max}$ condition at a lens mass of $M$, and then perform the calculation for $N_{exp}$ using the relevant $\tau_{FS}$ given $D_{max}$, the particular efficiency for the lens mass $\epsilon(M)$ and the number of stars within the bin. By iterating over all masses, we effectively calculate the number of expected events given the realistic distribution of stars in the AMPM catalogue. We calculate the number of expected events on a nightly basis over the duration of AMPM, and in doing so, can fold in the variation in the number of stars and the particular quirks in the detection efficiency of each night. The mean timescale $<t_{E}>$ is evaluated by taking the mean Einstein time of a subset of simulated events at every lens mass. 

\begin{figure}
\centering
\includegraphics[trim={0 0 0 0},clip,width=1.\columnwidth]{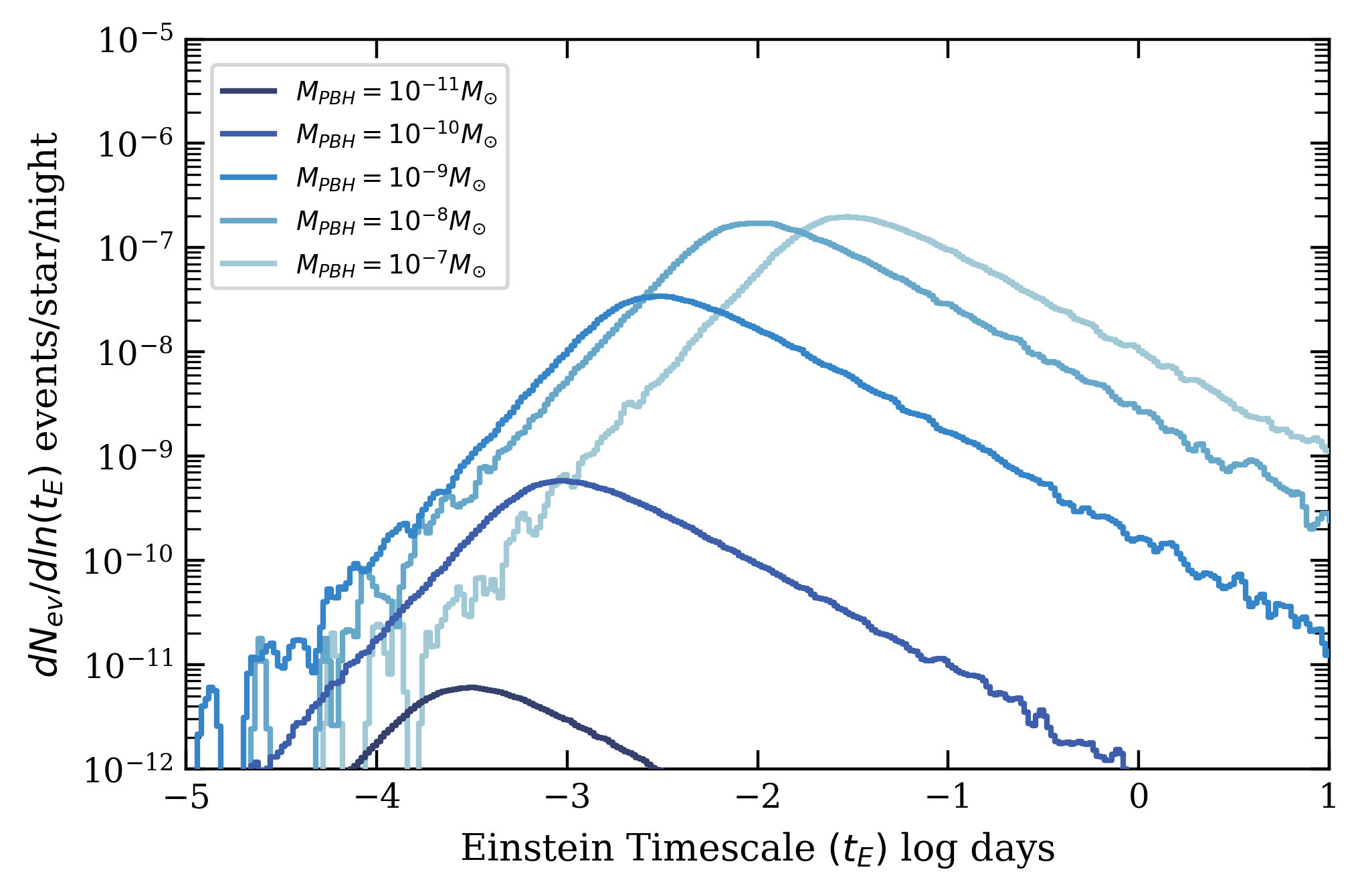}
\caption{The differential number of microlensing events by Einstein timescale ($t_E$) expected for the assumed NFW Milky Way dark halo at monochromatic mass steps for the PBH lens. Here, no efficiency weighting from the simulations is applied, such that $f =1$. }\label{fignightdiff}
\end{figure}

\begin{figure}
\centering
\includegraphics[trim={0 0 0 0},clip,width=1.\columnwidth]{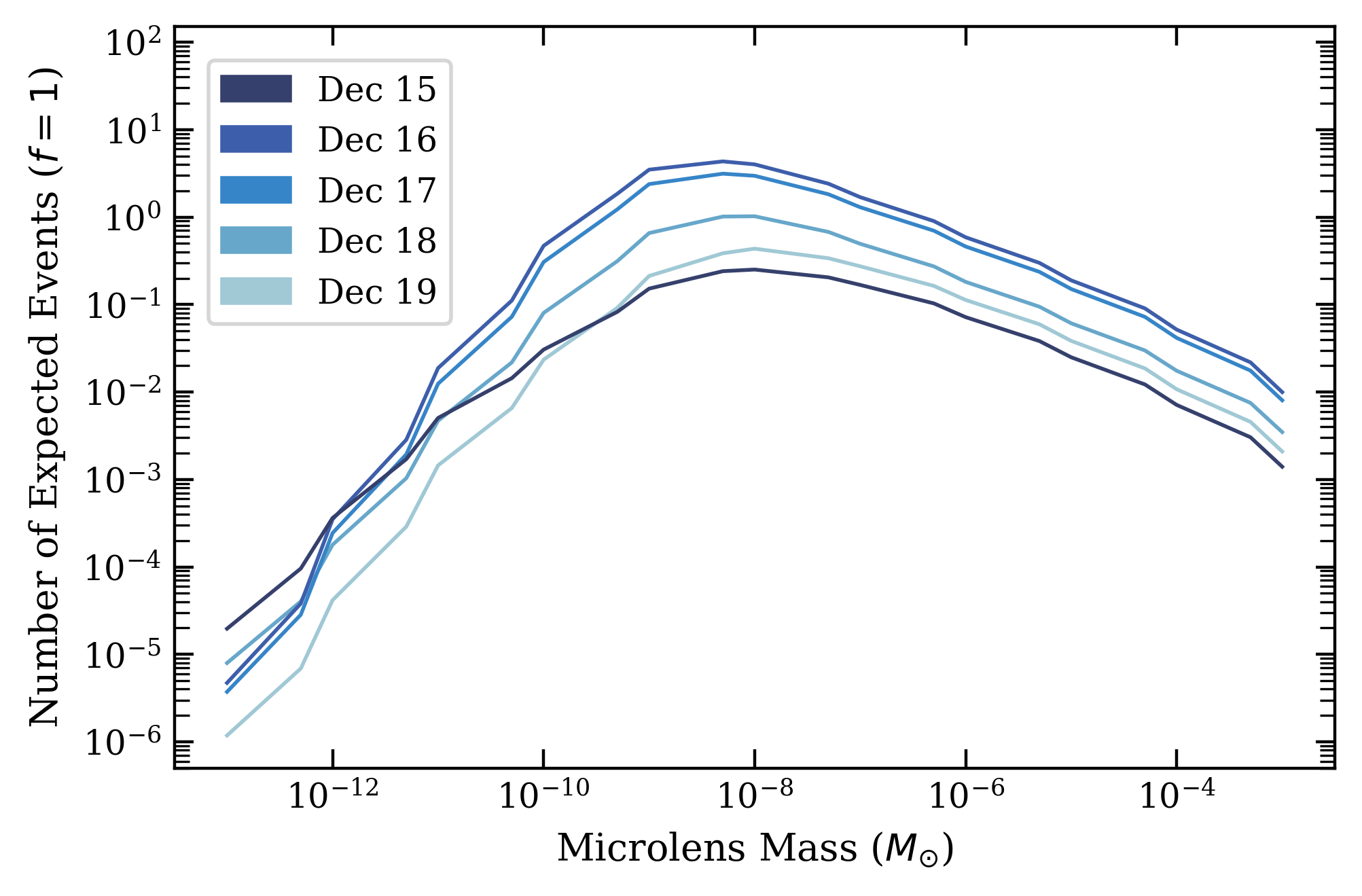}
\caption{The expected number of events per night in AMPM survey under the PS$-$PL and FS$-$PL treatments. The reduction to the optical depth sightline from the finite source damping effect is very evident at lens masses below $10^{-8}M_{\odot}$, at which point the rate of microlensing drops off steeply.}\label{fignightnexp}
\end{figure}

Figure \ref{fignightnexp} shows the number of expected events each night of the AMPM survey. The finite source $D_{max}$ threshold acts to reduce the amount of the dark halo that is open for investigation, where the lowest masses of PBHs must be located extremely close to the observer to produce visible microlensing. At masses below $10^{-10}M_{\odot}$, the NFW halo probed is essentially truncated to 0.05 kpc. Consequently, the rate of microlensing is strongly reduced, which is notably seen in Figure \ref{fignightnexp} as a turning point at the $10^{-9}M_{\odot}$ PBH mass, at which point, the rate of microlensing for smaller masses declines rapidly. 

The typical outcome of microlensing experiments has been to place 95\% confidence limit in the $M_\mathrm{PBH} - f_{DM}$ plane. The constraints are determined from the Poisson probability distribution for $n$ events detected from a smooth dark matter distribution with no background events \citep{Gorton_2022},
\begin{equation}
    P (n \,\, | \,N_{exp} )  = \frac{N^n_{exp}}{n!}e^{-N_{exp}}
\end{equation}
such that the 95\% confidence limit is defined by the value $x$, that gives $P(n = x) = 0.05$ \citep{Feldman_1998}.  To build the AMPM constraints on the dark matter fraction, $f$, we compare the 95\% confidence interval of $N_{exp}$, as defined in Equation \ref{eq:nexp}, to the number of events that should be detected given the totality of dark matter is contained in PBHs at any monochromatic mass. A null detection hypothesis of $n = 0$ at $P(n = 0)= 0.05$ requires $N_{exp} = 3$. Therefore, a scenario that predicts more than three PBH events at $f = 1$ can be ruled out to 95\% confidence. If we take our single microlensing detection to be a true PBH candidate, the $P(n = 1)= 0.05$ requires $N_{exp} = 4.74$ as the number of expected events.

\begin{figure}
\centering
\includegraphics[trim={0 0 0 0},clip,width=1.\columnwidth]{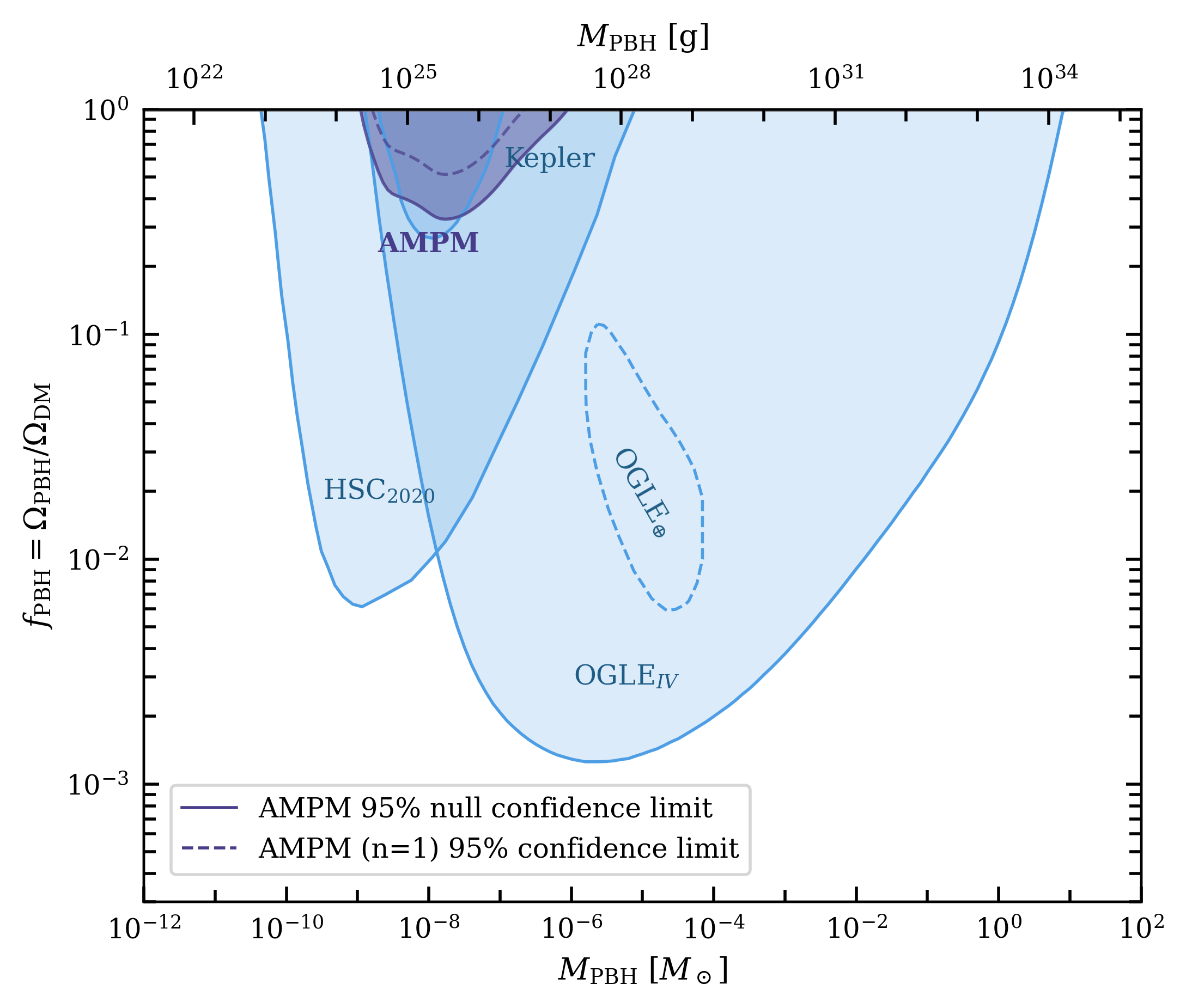}
\caption{The AMPM dark matter constraints evaluated at the 95\% upper bound on a null detection (solid dark purple line) and the 95\% confidence interval assuming a single detection (dashed shaded area). The relevant microlensing surveys for the AMPM mass sensitivity are shown for context, and compiled using the PBHBounds software \citep{PBHBounds}.}\label{figampmlimits}
\end{figure}

Figure \ref{figampmlimits} shows the outcome of AMPM and places our results in the context of the existing constraints on PBHs from the previous microlensing surveys. Shown as the solid line on Figure \ref{figampmlimits}, under the null detection hypothesis at the 95\% confidence limit, AMPM can constrain PBHs from comprising a fraction of the dark matter $f_{PBH} > 0.3$, with the strongest constraint at PBH mass of $10^{-8} M_{\odot}$. The dark matter constraints weaken to $f = 0.55$ by considering the single candidate detection. 

The microlensing constraint from each independent survey are derived from different sightlines and galactic modelling, for example AMPM only models the Milky Way dark matter, while Subaru-HSC and OGLE have combined the Milky Way with M31 or LMC to model the total distribution along their sightline. An increase in dark matter density equates to an increased microlensing rate, and hence stronger limits on the PBH-dark matter. Additionally, comparing the survey efficiencies is challenging given that any experimental design decisions, such as detection methods, sky coverage and data quality, become condensed into the single parameter. With some simplifying assumptions, and with all the caveats above, we can get an estimate of the consistency between AMPM and the Subaru-HSC and OGLE results by scaling from the exclusion limits in the $M_{PBH} - f_{DM}$ plane, as show in Figure \ref{figampmlimits} and with reference to Figure 8 in \citet{Sugiyama_2026}.

We compare the number of microlensing detections at a fiducial mass of $10^{-8} M_{\odot}$, which is the mass at which the AMPM,  Subaru-HSC and OGLE$IV$ limits are strongly constrained. At this mass, the constraining power differs between the AMPM and Subaru-HSC by a factor of 10, and by a factor of 30 for the OGLE$IV$ results. Naively, this would suggest that the rate of microlensing detections should scale according between each survey. Our single candidate is similar in timescale to the 12 Subaru-HSC candidates modelled with masses around $10^{-8} - 10^{-7} M_{\odot}$, meaning that the number of detections in AMPM and Subaru-HSC are in accordance given the survey sensitivities. However, AMPM's single PBH candidate suggests, when scaled to the constraining power of OGLE$IV$, that 30 candidates of similar hour-long timescales should have been detected by OGLE$IV$, where no detections were found \citep{Mroz_OGLEIV2024}.  

The results from AMPM would suggest that a far greater proportion of the dark matter content is comprised of PBHs than expected from OGLE$IV$. The possible resolutions for this apparent contradiction are;

\begin{itemize}\setlength{\itemsep}{3pt}
    \item[-] AMPM has overestimated the microlensing efficiency such that the true efficiency of our survey is much lower, and hence produces a lower microlensing candidate detection rate.
    \item[-] Conversely, OGLE$IV$, with their 15 minute cadence, has overestimated their efficiencies at the lower masses of $10^{-8} M_{\odot}$. In this case, multiple microlensing events may have remained unidentified in OGLE data. 
    \item[-] AMPM and Subaru-HSC has have detected a class of contaminating objects that are not PBH candidates, thus inflating their detection rates. It would be unlucky for both experiments to independently make these mistakes.
\end{itemize}

\section{Conclusions}

In this paper, we introduced a new microlensing survey, AMPM, targeting the LMC with five nights of DECam data in the wide VR-filter with one minute cadence. The rapid cadence of AMPM provides optimal resolution for very short-timescale microlensing events within the asteroid to planetary mass regime. We describe the observational program as well as the bespoke microlensing detection pipeline for short duration events. AMPM searches over two million quality light curves and found a single microlensing candidate with a FWHM timescale of $t_{FWHM} = 3.57$ hours. With our efficiency and Milky Way dark matter simulations, we find that at 95\% confidence limits under a null detection assumption, AMPM is able to constrain up to 30\% of the Milky Way dark matter as not consisting primarily of PBHs of monochromatic mass.

Looking forward, detecting microlensing PBHs with masses below $10^{-12} \, M_{\odot}$ is prohibited by finite source effects. The nearby stellar fields used in gravitational microlensing (LMC, SMC etc.) produce such damped and reduced microlensing signals that the detection of PBHs is unlikely. While not impossible, microlensing suffers strongly at the low mass lensing range to make the rate of PBH detection $<$ 1 per year. Imaging smaller stellar sources, such as white dwarfs \citep{Sugiyama_2020}, or more distant sources like those in M31 can mitigate the effects of finite source damping. However, sufficiently large numbers of extra-galactic white dwarfs are difficult to image at high cadences within a reasonable survey duration to place meaningful asteroid-mass constraints. Using M31 as a microlensing field for asteroid-mass PBHs requires rapid space-based observations to avoid the significant blending and atmospheric contamination rife in faint source light curves from ground-based observations. 

We do not say that microlensing as a PBH detection technique is no longer impactful for the detection of PBHs. On the contrary, we have a single compelling detection of an hours-long microlensing event, which we present the analysis and modelling of in the second paper of this series. The event is similar to the time scales of the candidate events detected from Subaru-HSC \citep{Niikura_2019, Sugiyama_2026}. With the combined detections, there is sufficient evidence for short-timescale microlenses attached to the Galactic dark halo. Whether these microlenses are baryonic (i.e. isolated moons/planets) or PBH dark matter requires further discussion and investigation. More detections can bolster the population statistics of the microlensing velocity, mass and distance distributions and aid in confirming the PBH nature of our detection. Therefore, we strongly advocate for additional microlensing experiments through galactic halo sightlines with specific interest in hour-long events. Such microlensing surveys are achievable with current detectors and should become a high priority for upcoming telescopes.

\section*{Acknowledgements}
This project used data obtained with the Dark Energy Camera (DECam), which was constructed by the Dark Energy Survey (DES) collaboration. Funding for the DES Projects has been provided by the US Department of Energy, the U.S. National Science Foundation, the Ministry of Science and Education of Spain, the Science and Technology Facilities Council of the United Kingdom, the Higher Education Funding Council for England, the National Center for Supercomputing Applications at the University of Illinois at Urbana-Champaign, the Kavli Institute for Cosmological Physics at the University of Chicago, Center for Cosmology and Astro-Particle Physics at the Ohio State University, the Mitchell Institute for Fundamental Physics and Astronomy at Texas A\&M University, Financiadora de Estudos e Projetos, Fundaç\"ao Carlos Chagas Filho de Amparo à Pesquisa do Estado do Rio de Janeiro, Conselho Nacional de Desenvolvimento Científico e Tecnol\'ogico and the Ministério da Ciência, Tecnologia e Inova\c{c}\"ao, the Deutsche Forschungsgemeinschaft and the Collaborating Institutions in the Dark Energy Survey.

The Collaborating Institutions are Argonne National Laboratory, the University of California at Santa Cruz, the University of Cambridge, Centro de Investigaciones Enérgeticas, Medioambientales y Tecnol\'ogicas–Madrid, the University of Chicago, University College London, the DES-Brazil Consortium, the University of Edinburgh, the Eidgen\"ossische Technische Hochschule (ETH) Z\"urich, Fermi National Accelerator Laboratory, the University of Illinois at Urbana-Champaign, the Institut de Ciències de l’Espai (IEEC/CSIC), the Institut de Física d’Altes Energies, Lawrence Berkeley National Laboratory, the Ludwig-Maximilians Universit\"at München and the associated Excellence Cluster Universe, the University of Michigan, NSF NOIRLab, the University of Nottingham, the Ohio State University, the OzDES Membership Consortium, the University of Pennsylvania, the University of Portsmouth, SLAC National Accelerator Laboratory, Stanford University, the University of Sussex, and Texas A\&M University.

Based on observations at NSF Cerro Tololo Inter-American Observatory, NSF NOIRLab (NOIRLab Prop. ID 2019B-0071; PI: J. Mould), which is managed by the Association of Universities for Research in Astronomy (AURA) under a cooperative agreement with the U.S. National Science Foundation.
This research has made use of "Aladin sky atlas" developed at CDS, Strasbourg Observatory, France. 

Computation was performed on the OzSTAR national facility at Swinburne University of Technology. The OzSTAR program receives funding in part from the Astronomy National Collaborative Research Infrastructure Strategy (NCRIS) allocation provided by the Australian Government, and from the Victorian Higher Education State Investment Fund (VHESIF) provided by the Victorian Government.

This research was supported by the Australian Research Council (ARC) Centre of Excellence for Dark Matter Particle Physics (CDM; centredarkmatter.org) with Grant Number: CE200100008. RK further acknowledges the support of the Australian Government Research Training Program (RTP) Scholarship doi.org/10.82133/C42F-K220

\section*{Data Availability}
The original data are available in the NOIRLab archive.
 



\bibliographystyle{mnras}
\bibliography{Ref} 

@article{DeRocco_2024,
   title={Revealing terrestrial-mass primordial black holes with the Nancy Grace Roman Space Telescope},
   volume={109},
   ISSN={2470-0029},
   url={http://dx.doi.org/10.1103/PhysRevD.109.023013},
   DOI={10.1103/physrevd.109.023013},
   number={2},
   journal={Physical Review D},
   publisher={American Physical Society (APS)},
   author={DeRocco, William and Frangipane, Evan and Hamer, Nick and Profumo, Stefano and Smyth, Nolan},
   year={2024},
   month=jan }

@misc{Green_2026,
      title={Stellar microlensing surveys as a probe of Primordial Black Holes: status and prospects}, 
      author={Anne M. Green},
      year={2026},
      eprint={2602.15974},
      archivePrefix={arXiv},
      primaryClass={astro-ph.GA},
      url={https://arxiv.org/abs/2602.15974}, 
}

@article{Lu_2019,
   title={Primordial Black Hole Microlensing: The Einstein Crossing Time Distribution},
   volume={3},
   ISSN={2515-5172},
   url={http://dx.doi.org/10.3847/2515-5172/ab1421},
   DOI={10.3847/2515-5172/ab1421},
   number={4},
   journal={Research Notes of the AAS},
   publisher={American Astronomical Society},
   author={Lu, Jessica R. and Lam, Casey Y. and Medford, Michael and Dawson, William and Golovich, Nathan},
   year={2019},
   month=apr, pages={58} }

@article{Poleski_2021, address={PL},
   title={Wide-Orbit Exoplanets are Common. Analysis of Nearly 20 Years of OGLE Microlensing Survey Data},
   volume={71},
   ISSN={00015237},
   url={https://doi.org/10.32023/0001-5237/71.1.1},
   DOI={10.32023/0001-5237/71.1.1},
   number={1},
   journal={Acta Astronomica},
   publisher={Copernicus Foundation for Polish Astronomy},
   author={Poleski, R. and Skowron, J. and Mróz, P. and Udalski, A. and Szymański, M.K. and Pietrukowicz, P. and Ulaczyk, K. and Rybicki, K. and Iwanek, P. and Wrona, M. and Gromadzki, M.},
   year={2021},
   month=Mar, pages={1–23} }

@article{Zang_2025,
   title={Microlensing events indicate that super-Earth exoplanets are common in Jupiter-like orbits},
   volume={388},
   ISSN={1095-9203},
   url={http://dx.doi.org/10.1126/science.adn6088},
   DOI={10.1126/science.adn6088},
   number={6745},
   journal={Science},
   publisher={American Association for the Advancement of Science (AAAS)},
   author={Zang, Weicheng and Jung, Youn Kil and Yee, Jennifer C. and Hwang, Kyu-Ha and Yang, Hongjing and Udalski, Andrzej and Sumi, Takahiro and Gould, Andrew and Mao, Shude and Albrow, Michael D. and Chung, Sun-Ju and Han, Cheongho and Ryu, Yoon-Hyun and Shin, In-Gu and Shvartzvald, Yossi and Cha, Sang-Mok and Kim, Dong-Jin and Kim, Hyoun-Woo and Kim, Seung-Lee and Lee, Chung-Uk and Lee, Dong-Joo and Lee, Yongseok and Park, Byeong-Gon and Pogge, Richard W. and Zhang, Xiangyu and Kuang, Renkun and Wang, Hanyue and Zhang, Jiyuan and Hu, Zhecheng and Zhu, Wei and Mróz, Przemek and Skowron, Jan and Poleski, Radosław and Szymański, Michał K. and Soszyński, Igor and Pietrukowicz, Paweł and Kozłowski, Szymon and Ulaczyk, Krzysztof and Rybicki, Krzysztof A. and Iwanek, Patryk and Wrona, Marcin and Gromadzki, Mariusz and Abe, Fumio and Barry, Richard and Bennett, David P. and Bhattacharya, Aparna and Bond, Ian A. and Fujii, Hirosane and Fukui, Akihiko and Hamada, Ryusei and Hirao, Yuki and Silva, Stela Ishitani and Itow, Yoshitaka and Kirikawa, Rintaro and Koshimoto, Naoki and Matsubara, Yutaka and Miyazaki, Shota and Muraki, Yasushi and Olmschenk, Greg and Ranc, Clément and Rattenbury, Nicholas J. and Satoh, Yuki and Suzuki, Daisuke and Tomoyoshi, Mio and Tristram, Paul J. and Vandorou, Aikaterini and Yama, Hibiki and Yamashita, Kansuke},
   year={2025},
   month=Apr, pages={400–404} }

@ARTICLE{Han_2005,
       author = {{Han}, Cheongho and {Gaudi}, B. Scott and {An}, Jin H. and {Gould}, Andrew},
        title = "{Microlensing Detection and Characterization of Wide-Separation Planets}",
      journal = {\apj},
         year = 2005,
        month = jan,
       volume = {618},
       number = {2},
        pages = {962-972},
          doi = {10.1086/426115},
archivePrefix = {arXiv},
       eprint = {astro-ph/0409589},
 primaryClass = {astro-ph},
       adsurl = {https://ui.adsabs.harvard.edu/abs/2005ApJ...618..962H}
}

@article{Han_2004,
doi = {10.1086/381429},
url = {https://doi.org/10.1086/381429},
year = {2004},
month = {mar},
publisher = {},
volume = {604},
number = {1},
pages = {372},
author = {Han, Cheongho and Chung, Sun-Ju and Kim, Doeon and Park, Byeong-Gon and Ryu, Yoon-Hyun and Kang, Sangjun and Lee, Dong Wook},
title = {Gravitational Microlensing: A Tool for Detecting and Characterizing Free-Floating Planets},
journal = {The Astrophysical Journal},
}

@article{Dong_2026,
   title={A free-floating-planet microlensing event caused by a Saturn-mass object},
   volume={391},
   ISSN={1095-9203},
   url={http://dx.doi.org/10.1126/science.adv9266},
   DOI={10.1126/science.adv9266},
   number={6780},
   journal={Science},
   publisher={American Association for the Advancement of Science (AAAS)},
   author={Dong, Subo and Wu, Zexuan and Ryu, Yoon-Hyun and Udalski, Andrzej and Mróz, Przemek and Rybicki, Krzysztof A. and Hodgkin, Simon T. and Wyrzykowski, Łukasz and Eyer, Laurent and Bensby, Thomas and Chen, Ping and Wang, Sharon X. and Gould, Andrew and Yang, Hongjing and Albrow, Michael D. and Chung, Sun-Ju and Han, Cheongho and Hwang, Kyu-Ha and Jung, Youn Kil and Shin, In-Gu and Shvartzvald, Yossi and Yee, Jennifer C. and Zang, Weicheng and Kim, Dong-Jin and Lee, Chung-Uk and Park, Byeong-Gon and Poleski, Radosław and Skowron, Jan and Szymański, Michał K. and Soszyński, Igor and Pietrukowicz, Paweł and Kozłowski, Szymon and Skowron, Dorota M. and Ulaczyk, Krzysztof and Gromadzki, Mariusz and Ratajczak, Milena and Iwanek, Patryk and Wrona, Marcin and Mróz, Mateusz J. and Rixon, Guy and Harrison, Diana L. and Breedt, Elmé},
   year={2026},
   month=Jan, pages={96–99} }

@Inbook{Carr_2025,
author="Carr, Bernard J.
and Green, Anne M.",
title="The History of Primordial Black Holes",
bookTitle="Primordial Black Holes",
year="2025",
publisher="Springer Nature Singapore",
address="Singapore",
pages="3--33",
isbn="978-981-97-8887-3",
doi="10.1007/978-981-97-8887-3_1",
url="https://doi.org/10.1007/978-981-97-8887-3_1"
}

@ARTICLE{Sumi_2011,
       author = {{Sumi}, T. and {Kamiya}, K. and {Bennett}, D.~P. and {Bond}, I.~A. and {Abe}, F. and {Botzler}, C.~S. and {Fukui}, A. and {Furusawa}, K. and {Hearnshaw}, J.~B. and {Itow}, Y. and {Kilmartin}, P.~M. and {Korpela}, A. and {Lin}, W. and {Ling}, C.~H. and {Masuda}, K. and {Matsubara}, Y. and {Miyake}, N. and {Motomura}, M. and {Muraki}, Y. and {Nagaya}, M. and {Nakamura}, S. and {Ohnishi}, K. and {Okumura}, T. and {Perrott}, Y.~C. and {Rattenbury}, N. and {Saito}, To. and {Sako}, T. and {Sullivan}, D.~J. and {Sweatman}, W.~L. and {Tristram}, P.~J. and {Udalski}, A. and {Szyma{\'n}ski}, M.~K. and {Kubiak}, M. and {Pietrzy{\'n}ski}, G. and {Poleski}, R. and {Soszy{\'n}ski}, I. and {Wyrzykowski}, {\L}. and {Ulaczyk}, K. and {Microlensing Observations in Astrophysics (MOA) Collaboration}},
        title = "{Unbound or distant planetary mass population detected by gravitational microlensing}",
      journal = {\nat},
         year = 2011,
        month = may,
       volume = {473},
       number = {7347},
        pages = {349-352},
          doi = {10.1038/nature10092},
archivePrefix = {arXiv},
       eprint = {1105.3544},
 primaryClass = {astro-ph.EP},
       adsurl = {https://ui.adsabs.harvard.edu/abs/2011Natur.473..349S}
}

@ARTICLE{Ju_2022,
author = {{Chung}, Sun-Ju and {Yee}, Jennifer C. and {Udalski}, Andrej and {Gould}, Andrew and {Albrow}, Michael D. and {Jung}, Youn Kil and {Hwang}, Kyu-Ha and {Han}, Cheongho and {Ryu}, Yoon-Hyun and {Shin}, In-Gu and {Shvartzvald}, Yossi and {Zang}, Weicheng and {Cha}, Sang-Mok and {Kim}, Dong-Jin and {Kim}, Seung-Lee and {Lee}, Chung-Uk and {Lee}, Dong-Joo and {Lee}, Yongseok and {Park}, Byeong-Gon and {Pogge}, Richard W. and {Poleski}, Radek and {Mr{\'o}z}, Przemek and {Pietrukowicz}, Pawe{\l} and {Skowron}, Jan and {Szyma{\'n}ski}, Micha{\l} K. and {Soszy{\'n}ski}, Igor and {Koz{\l}owski}, Szymon and {Rybicki}, Krzysztof A. and {Iwanek}, Patryk and {Wrona}, Marcin and {Gromadzki}, Mariusz and {Ulaczyk}, Krzysztof},
title = "{OGLE-2019-BLG-0362Lb: A Super-Jovian-Mass Planet around a Low-Mass Star}",
journal = {Journal of Korean Astronomical Society},
year = 2022,
month = aug,
volume = {55},
pages = {123-130},
doi = {10.5303/JKAS.2022.55.4.123},
archivePrefix = {arXiv},
eprint = {2208.04230},
primaryClass = {astro-ph.EP},
adsurl = {https://ui.adsabs.harvard.edu/abs/2022JKAS...55..123C}}

@ARTICLE{Beaulieu_2006,
       author = {{Beaulieu}, J.-P. and {Bennett}, D.~P. and {Fouqu{\'e}}, P. and {Williams}, A. and {Dominik}, M. and {J{\o}rgensen}, U.~G. and {Kubas}, D. and {Cassan}, A. and {Coutures}, C. and {Greenhill}, J. and {Hill}, K. and {Menzies}, J. and {Sackett}, P.~D. and {Albrow}, M. and {Brillant}, S. and {Caldwell}, J.~A.~R. and {Calitz}, J.~J. and {Cook}, K.~H. and {Corrales}, E. and {Desort}, M. and {Dieters}, S. and {Dominis}, D. and {Donatowicz}, J. and {Hoffman}, M. and {Kane}, S. and {Marquette}, J.-B. and {Martin}, R. and {Meintjes}, P. and {Pollard}, K. and {Sahu}, K. and {Vinter}, C. and {Wambsganss}, J. and {Woller}, K. and {Horne}, K. and {Steele}, I. and {Bramich}, D.~M. and {Burgdorf}, M. and {Snodgrass}, C. and {Bode}, M. and {Udalski}, A. and {Szyma{\'n}ski}, M.~K. and {Kubiak}, M. and {Wi{\c{e}}ckowski}, T. and {Pietrzy{\'n}ski}, G. and {Soszy{\'n}ski}, I. and {Szewczyk}, O. and {Wyrzykowski}, {\L}. and {Paczy{\'n}ski}, B. and {Abe}, F. and {Bond}, I.~A. and {Britton}, T.~R. and {Gilmore}, A.~C. and {Hearnshaw}, J.~B. and {Itow}, Y. and {Kamiya}, K. and {Kilmartin}, P.~M. and {Korpela}, A.~V. and {Masuda}, K. and {Matsubara}, Y. and {Motomura}, M. and {Muraki}, Y. and {Nakamura}, S. and {Okada}, C. and {Ohnishi}, K. and {Rattenbury}, N.~J. and {Sako}, T. and {Sato}, S. and {Sasaki}, M. and {Sekiguchi}, T. and {Sullivan}, D.~J. and {Tristram}, P.~J. and {Yock}, P.~C.~M. and {Yoshioka}, T.},
        title = "{Discovery of a cool planet of 5.5 Earth masses through gravitational microlensing}",
      journal = {\nat},
         year = 2006,
        month = jan,
       volume = {439},
       number = {7075},
        pages = {437-440},
          doi = {10.1038/nature04441},
archivePrefix = {arXiv},
       eprint = {astro-ph/0601563},
 primaryClass = {astro-ph},
       adsurl = {https://ui.adsabs.harvard.edu/abs/2006Natur.439..437B}
}

@article{Mroz_2020_terrestrial,
doi = {10.3847/2041-8213/abbfad},
url = {https://doi.org/10.3847/2041-8213/abbfad},
year = {2020},
month = {oct},
publisher = {The American Astronomical Society},
volume = {903},
number = {1},
pages = {L11},
author = {Mróz, Przemek and Poleski, Radosław and Gould, Andrew and Udalski, Andrzej and Sumi, Takahiro and and and Szymański, Michał K. and Soszyński, Igor and Pietrukowicz, Paweł and Kozłowski, Szymon and Skowron, Jan and Ulaczyk, Krzysztof and (OGLE Collaboration) and Albrow, Michael D. and Chung, Sun-Ju and Han, Cheongho and Hwang, Kyu-Ha and Jung, Youn Kil and Kim, Hyoun-Woo and Ryu, Yoon-Hyun and Shin, In-Gu and Shvartzvald, Yossi and Yee, Jennifer C. and Zang, Weicheng and Cha, Sang-Mok and Kim, Dong-Jin and Kim, Seung-Lee and Lee, Chung-Uk and Lee, Dong-Joo and Lee, Yongseok and Park, Byeong-Gon and Pogge, Richard W. and (KMT Collaboration)},
title = {A Terrestrial-mass Rogue Planet Candidate Detected in the Shortest-timescale Microlensing Event},
journal = {The Astrophysical Journal Letters},
}

@ARTICLE{Carr_1990,
       author = {{Carr}, Bernard and {Primack}, Joel},
        title = "{Searching for MACHOs}",
      journal = {\nat},
         year = 1990,
        month = jun,
       volume = {345},
       number = {6275},
        pages = {478-479},
          doi = {10.1038/345478a0},
       adsurl = {https://ui.adsabs.harvard.edu/abs/1990Natur.345..478C}
}

@ARTICLE{Nunota_2025,
       author = {{Nunota}, Kansuke and {Sumi}, Takahiro and {Koshimoto}, Naoki and {Rattenbury}, Nicholas J. and {Abe}, Fumio and {Barry}, Richard and {Bennett}, David P. and {Bhattacharya}, Aparna and {Fukui}, Akihiko and {Hamada}, Ryusei and {Hamada}, Shunya and {Hamasaki}, Naoto and {Hirao}, Yuki and {Ishitani Silva}, Stela and {Itow}, Yoshitaka and {Matsubara}, Yutaka and {Miyazaki}, Shota and {Muraki}, Yasushi and {Nagai}, Tsutsumi and {Olmschenk}, Greg and {Ranc}, Clement and {Satoh}, Yuki K. and {Suzuki}, Daisuke and {Tristram}, Paul. J. and {Vandorou}, Aikaterini and {Yama}, Hibiki and {MOA Collaboration}},
        title = "{The Microlensing Event Rate and Optical Depth from MOA-II 9 Yr Survey Toward the Galactic Bulge}",
      journal = {\apj},
         year = 2025,
        month = feb,
       volume = {979},
       number = {2},
          eid = {123},
        pages = {123},
          doi = {10.3847/1538-4357/ada352},
archivePrefix = {arXiv},
       eprint = {2410.23553},
}

@article{Alcock_2001,
   title={MACHO Project Limits on Black Hole Dark Matter in the 1–30 [ITAL]M[/ITAL][TINF]⊙[/TINF] Range},
   volume={550},
   ISSN={0004-637X},
   url={http://dx.doi.org/10.1086/319636},
   DOI={10.1086/319636},
   number={2},
   journal={The Astrophysical Journal},
   publisher={American Astronomical Society},
   author={Alcock, C. and Allsman, R. A. and Alves, D. R. and Axelrod, T. S. and Becker, A. C. and Bennett, D. P. and Cook, K. H. and Dalal, N. and Drake, A. J. and Freeman, K. C. and Geha, M. and Griest, K. and Lehner, M. J. and Marshall, S. L. and Minniti, D. and Nelson, C. A. and Peterson, B. A. and Popowski, P. and Pratt, M. R. and Quinn, P. J. and Stubbs, C. W. and Sutherland, W. and Tomaney, A. B. and Vandehei, T. and Welch, D. L. and (The MACHO Collaboration)},
   year={2001},
   month=apr, pages={L169–L172} }

@article{Alcock_2000,
   title={The MACHO Project: Microlensing Results from 5.7 Years of Large Magellanic Cloud Observations},
   volume={542},
   ISSN={1538-4357},
   DOI={10.1086/309512},
   number={1},
   journal={The Astrophysical Journal},
   publisher={American Astronomical Society},
   author={Alcock, C. and Allsman, R. A. and Alves, D. R. and Axelrod, T. S. and Becker, A. C. and Bennett, D. P. and Cook, K. H. and Dalal, N. and Drake, A. J. and Freeman, K. C. and Geha, M. and Griest, K. and Lehner, M. J. and Marshall, S. L. and Minniti, D. and Nelson, C. A. and Peterson, B. A. and Popowski, P. and Pratt, M. R. and Quinn, P. J. and Stubbs, C. W. and Sutherland, W. and Tomaney, A. B. and Vandehei, T. and Welch, D.},
   year={2000},
   month=oct, pages={281–307} }

@article{Kunimoto_2025,
doi = {10.3847/1538-3881/ae10ab},
url = {https://doi.org/10.3847/1538-3881/ae10ab},
year = {2025},
month = {nov},
publisher = {The American Astronomical Society},
volume = {170},
number = {6},
pages = {321},
author = {Kunimoto, Michelle and DeRocco, William and Smyth, Nolan and Bryson, Steve and Gaudi, B. Scott},
title = {Searching for Free-floating Planets with TESS: Results from Sectors 61–65},
journal = {The Astronomical Journal},
}

@article{DeRocco_2023,
   title={Constraints on sub-terrestrial free-floating planets from Subaru microlensing observations},
   volume={527},
   ISSN={1365-2966},
   DOI={10.1093/mnras/stad3824},
   number={3},
   journal={Monthly Notices of the Royal Astronomical Society},
   publisher={Oxford University Press (OUP)},
   author={DeRocco, William and Smyth, Nolan and Profumo, Stefano},
   year={2023},
   month=nov, pages={8921–8930} }

@article{Mroz_OGLEIV2024,
   title={Limits on Planetary-mass Primordial Black Holes from the OGLE High-cadence Survey of the Magellanic Clouds},
   volume={976},
   ISSN={2041-8213},
   DOI={10.3847/2041-8213/ad8e68},
   number={1},
   journal={The Astrophysical Journal Letters},
   publisher={American Astronomical Society},
   author={Mr\'oz, Przemek and Udalski, Andrzej and Szymański, Michał K. and Soszyński, Igor and Pietrukowicz, Paweł and Kozłowski, Szymon and Poleski, Radosław and Skowron, Jan and Ulaczyk, Krzysztof and Gromadzki, Mariusz and Rybicki, Krzysztof and Iwanek, Patryk and Wrona, Marcin and Mróz, Mateusz J.},
   year={2024},
   month=nov, pages={L19} }

@article{Bird_2016,
   title={Did LIGO Detect Dark Matter?},
   volume={116},
   ISSN={1079-7114},
   url={http://dx.doi.org/10.1103/PhysRevLett.116.201301},
   DOI={10.1103/physrevlett.116.201301},
   number={20},
   journal={Physical Review Letters},
   publisher={American Physical Society (APS)},
   author={Bird, Simeon and Cholis, Ilias and Muñoz, Julian B. and Ali-Haïmoud, Yacine and Kamionkowski, Marc and Kovetz, Ely D. and Raccanelli, Alvise and Riess, Adam G.},
   year={2016},
   month=may }

@article{Salucci_2019,
   title={The distribution of dark matter in galaxies},
   volume={27},
   ISSN={1432-0754},
   url={http://dx.doi.org/10.1007/s00159-018-0113-1},
   DOI={10.1007/s00159-018-0113-1},
   number={1},
   journal={The Astronomy and Astrophysics Review},
   publisher={Springer Science and Business Media LLC},
   author={Salucci, Paolo},
   year={2019},
   month=feb }

@article{Clesse_2022,
title = {GW190425, GW190521 and GW190814: Three candidate mergers of primordial black holes from the QCD epoch},
journal = {Physics of the Dark Universe},
volume = {38},
pages = {101111},
year = {2022},
issn = {2212-6864},
doi = {https://doi.org/10.1016/j.dark.2022.101111},
url = {https://www.sciencedirect.com/science/article/pii/S221268642200084X},
author = {Sebastien Clesse and Juan García-Bellido}
}

@article{Liu_2022,
    author = {Liu, Boyuan and Zhang, Saiyang and Bromm, Volker},
    title = {Effects of stellar-mass primordial black holes on first star formation},
    journal = {Monthly Notices of the Royal Astronomical Society},
    volume = {514},
    number = {2},
    pages = {2376-2396},
    year = {2022},
    month = {05},
    issn = {0035-8711},
    doi = {10.1093/mnras/stac1472},
    url = {https://doi.org/10.1093/mnras/stac1472},
    eprint = {https://academic.oup.com/mnras/article-pdf/514/2/2376/44114702/stac1472.pdf},
}

@article{Dayal_2024,
   title={Exploring a primordial solution for early black holes detected with JWST},
   volume={690},
   ISSN={1432-0746},
   url={http://dx.doi.org/10.1051/0004-6361/202451481},
   DOI={10.1051/0004-6361/202451481},
   journal={Astronomy & Astrophysics},
   publisher={EDP Sciences},
   author={Dayal, Pratika},
   year={2024},
   month=oct, pages={A182} }

@ARTICLE{Colazo_2024,
       author = {{Colazo}, P.~E. and {Stasyszyn}, F. and {Padilla}, N.},
        title = "{Structure formation with primordial black holes to alleviate early star formation tension revealed by JWST}",
      journal = {\aap},
         year = 2024,
        month = may,
       volume = {685},
          eid = {L8},
        pages = {L8},
          doi = {10.1051/0004-6361/202449565},
archivePrefix = {arXiv},
       eprint = {2404.13110},
 primaryClass = {astro-ph.CO},
       adsurl = {https://ui.adsabs.harvard.edu/abs/2024A&A...685L...8C}
}

@article{Flores_2023,
  title = {Structure formation after reheating: Supermassive primordial black holes and Fermi ball dark matter},
  author = {Flores, Marcos M. and Lu, Yifan and Kusenko, Alexander},
  journal = {Phys. Rev. D},
  volume = {108},
  issue = {12},
  pages = {123511},
  numpages = {10},
  year = {2023},
  month = {Dec},
  publisher = {American Physical Society},
  doi = {10.1103/PhysRevD.108.123511},
  url = {https://link.aps.org/doi/10.1103/PhysRevD.108.123511}
}

@article{Ryan_2022,
  title = {Exotic compact objects: The dark white dwarf},
  author = {Ryan, Michael and Radice, David},
  journal = {Phys. Rev. D},
  volume = {105},
  issue = {11},
  pages = {115034},
  numpages = {12},
  year = {2022},
  month = {Jun},
  publisher = {American Physical Society},
  doi = {10.1103/PhysRevD.105.115034},
  url = {https://link.aps.org/doi/10.1103/PhysRevD.105.115034}
}

@article{Fujikura,
  title = {Microlensing constraints on axion stars including finite lens and source size effects},
  author = {Fujikura, Kohei and Hertzberg, Mark P. and Schiappacasse, Enrico D. and Yamaguchi, Masahide},
  journal = {Phys. Rev. D},
  volume = {104},
  issue = {12},
  pages = {123012},
  numpages = {20},
  year = {2021},
  month = {Dec},
  publisher = {American Physical Society},
  doi = {10.1103/PhysRevD.104.123012},
  url = {https://link.aps.org/doi/10.1103/PhysRevD.104.123012}
}

@article{Villanueva_Domingo_2021,
   title={A Brief Review on Primordial Black Holes as Dark Matter},
   volume={8},
   ISSN={2296-987X},
   url={http://dx.doi.org/10.3389/fspas.2021.681084},
   DOI={10.3389/fspas.2021.681084},
   journal={Frontiers in Astronomy and Space Sciences},
   publisher={Frontiers Media SA},
   author={Villanueva-Domingo, Pablo and Mena, Olga and Palomares-Ruiz, Sergio},
   year={2021},
   month=may }

@article{Carr_2024,
   title={Observational evidence for primordial black holes: A positivist perspective},
   volume={1054},
   ISSN={0370-1573},
   url={http://dx.doi.org/10.1016/j.physrep.2023.11.005},
   DOI={10.1016/j.physrep.2023.11.005},
   journal={Physics Reports},
   publisher={Elsevier BV},
   author={Carr, B.J. and Clesse, S. and García-Bellido, J. and Hawkins, M.R.S. and Kühnel, F.},
   year={2024},
   month=feb, pages={1–68} }

@misc{Sugiyama_2026,
      title={Microlensing constraints on Primordial Black Hole abundance with Subaru Hyper Suprime-Cam observations of Andromeda}, 
      author={Sunao Sugiyama and Masahiro Takada and Naoki Yasuda and Nozomu Tominaga},
      year={2026},
      eprint={2602.05840},
      archivePrefix={arXiv},
      primaryClass={astro-ph.CO},
      url={https://arxiv.org/abs/2602.05840}, 
}

@article{Mroz_2017,
   title={No large population of unbound or wide-orbit Jupiter-mass planets},
   volume={548},
   ISSN={1476-4687},
   url={http://dx.doi.org/10.1038/nature23276},
   DOI={10.1038/nature23276},
   number={7666},
   journal={Nature},
   publisher={Springer Science and Business Media LLC},
   author={Mr\'oz, Przemek and Udalski, Andrzej and Skowron, Jan and Poleski, Radosław and Kozłowski, Szymon and Szymański, Michał K. and Soszyński, Igor and Wyrzykowski, Łukasz and Pietrukowicz, Paweł and Ulaczyk, Krzysztof and Skowron, Dorota and Pawlak, Michał},
   year={2017},
   month=jul, pages={183–186} }

@article{Mroz_2019,
   title={Two new free-floating or wide-orbit planets from microlensing},
   volume={622},
   ISSN={1432-0746},
   url={http://dx.doi.org/10.1051/0004-6361/201834557},
   DOI={10.1051/0004-6361/201834557},
   journal={Astronomy &amp; Astrophysics},
   publisher={EDP Sciences},
   author={Mróz, Przemek and Udalski, Andrzej and Bennett, David P. and Ryu, Yoon-Hyun and Sumi, Takahiro and Shvartzvald, Yossi and Skowron, Jan and Poleski, Radosław and Pietrukowicz, Paweł and Kozłowski, Szymon and Szymański, Michał K. and Wyrzykowski, Łukasz and Soszyński, Igor and Ulaczyk, Krzysztof and Rybicki, Krzysztof and Iwanek, Patryk and Albrow, Michael D. and Chung, Sun-Ju and Gould, Andrew and Han, Cheongho and Hwang, Kyu-Ha and Jung, Youn Kil and Shin, In-Gu and Yee, Jennifer C. and Zang, Weicheng and Cha, Sang-Mok and Kim, Dong-Jin and Kim, Hyoun-Woo and Kim, Seung-Lee and Lee, Chung-Uk and Lee, Dong-Joo and Lee, Yongseok and Park, Byeong-Gon and Pogge, Richard W. and Abe, Fumio and Barry, Richard and Bhattacharya, Aparna and Bond, Ian A. and Donachie, Martin and Fukui, Akihiko and Hirao, Yuki and Itow, Yoshitaka and Kawasaki, Kohei and Kondo, Iona and Koshimoto, Naoki and Li, Man Cheung Alex and Matsubara, Yutaka and Muraki, Yasushi and Miyazaki, Shota and Nagakane, Masayuki and Ranc, Clément and Rattenbury, Nicholas J. and Suematsu, Haruno and Sullivan, Denis J. and Suzuki, Daisuke and Tristram, Paul J. and Yonehara, Atsunori and Maoz, Dan and Kaspi, Shai and Friedmann, Matan},
   year={2019},
   month=Feb, pages={A201} }

@article{Niikura_OGLE2019,
   title={Constraints on Earth-mass primordial black holes from OGLE 5-year microlensing events},
   volume={99},
   ISSN={2470-0029},
   url={http://dx.doi.org/10.1103/PhysRevD.99.083503},
   DOI={10.1103/physrevd.99.083503},
   number={8},
   journal={Physical Review D},
   publisher={American Physical Society (APS)},
   author={Niikura, Hiroko and Takada, Masahiro and Yokoyama, Shuichiro and Sumi, Takahiro and Masaki, Shogo},
   year={2019},
   month=apr }

@article{Niikura_2019,
   title={Microlensing constraints on primordial black holes with Subaru/HSC Andromeda observations},
   volume={3},
   ISSN={2397-3366},
   DOI={10.1038/s41550-019-0723-1},
   number={6},
   journal={Nature Astronomy},
   publisher={Springer Science and Business Media LLC},
   author={Niikura, Hiroko and Takada, Masahiro and Yasuda, Naoki and Lupton, Robert H. and Sumi, Takahiro and More, Surhud and Kurita, Toshiki and Sugiyama, Sunao and More, Anupreeta and Oguri, Masamune and Chiba, Masashi},
   year={2019},
   month=apr, pages={524–534} }

@article{Green_2024,
title = {Primordial black holes as a dark matter candidate - a brief overview},
journal = {Nuclear Physics B},
volume = {1003},
pages = {116494},
year = {2024},
note = {Special Issue of Nobel Symposium 182 on Dark Matter},
issn = {0550-3213},
doi = {https://doi.org/10.1016/j.nuclphysb.2024.116494},
author = {Anne M. Green}}

@article{Paczyski_1996,
   title={GRAVITATIONAL MICROLENSING IN THE LOCAL GROUP},
   volume={34},
   ISSN={1545-4282},
   DOI={10.1146/annurev.astro.34.1.419},
   number={1},
   journal={Annual Review of Astronomy and Astrophysics},
   publisher={Annual Reviews},
   author={Paczy\'nski, Bohdan},
   year={1996},
   month=sep, pages={419–459} }

@ARTICLE{Paczynski_1986,
       author = {{Paczy\'nski}, B.},
        title = "{Gravitational Microlensing by the Galactic Halo}",
      journal = {\apj},
         year = 1986,
        month = may,
       volume = {304},
        pages = {1},
          doi = {10.1086/164140},}

@book{Schneider_1992,
author = {Schneider, Peter and Ehlers, Jürgen and Falco, Emilio E},
address = {Berlin, Heidelberg},
copyright = {Springer-Verlag Berlin Heidelberg 1992},
edition = {1},
isbn = {1461276551},
issn = {0941-7834},
language = {eng},
organization = {SpringerLink (Online service)},
publisher = {Springer},
series = {Astronomy and Astrophysics Library},
title = {Gravitational lenses},
year = {1992},
}

@ARTICLE{WittMao_1994,
       author = {{Witt}, Hans J. and {Mao}, Shude},
        title = "{Can Lensed Stars Be Regarded as Pointlike for Microlensing by MACHOs?}",
      journal = {\apj},
         year = 1994,
        month = aug,
       volume = {430},
        pages = {505},
          doi = {10.1086/174426},
}

@ARTICLE{Griest_1991,
       author = {{Griest}, Kim},
        title = "{Galactic Microlensing as a Method of Detecting Massive Compact Halo Objects}",
      journal = {\apj},
         year = 1991,
        month = jan,
       volume = {366},
        pages = {412},
          doi = {10.1086/169575},}

@article{Griest_2011,
   title={Microlensing of Kepler Stars as a Method of Detecting Primordial Black Hole Dark Matter},
   volume={107},
   ISSN={1079-7114},
   DOI={10.1103/physrevlett.107.231101},
   number={23},
   journal={Physical Review Letters},
   publisher={American Physical Society (APS)},
   author={Griest, Kim and Lehner, Matthew J. and Cieplak, Agnieszka M. and Jain, Bhuvnesh},
   year={2011},
   month=dec }

@ARTICLE{Lee_2009,
       author = {{Lee}, C. -H. and {Riffeser}, A. and {Seitz}, S. and {Bender}, R.},
        title = "{Finite-Source Effects in Microlensing: A Precise, Easy to Implement, Fast, and Numerically Stable Formalism}",
      journal = {\apj},
         year = 2009,
        month = apr,
       volume = {695},
       number = {1},
        pages = {200-207},
          doi = {10.1088/0004-637X/695/1/200}}

@ARTICLE{NO_1984,
       author = {{Nityananda}, R. and {Ostriker}, J.~P.},
        title = "{Gravitational Lensing by Stars in a Galaxy Halo - Theory of Combined Weak and Strong Scattering}",
      journal = {Journal of Astrophysics and Astronomy},
         year = 1984,
        month = sep,
       volume = {5},
       number = {3},
        pages = {235-250},
          doi = {10.1007/BF02714541}}

@ARTICLE{VO_1983,
       author = {{Vietri}, M. and {Ostriker}, J.~P.},
        title = "{The statistics of gravitational lenses - Appaarent changes in the luminosity function of distant sources due to passage of light through a single galaxy}",
      journal = {\apj},
         year = 1983,
        month = apr,
       volume = {267},
        pages = {488-510},
          doi = {10.1086/160886}}

@ARTICLE{Cieplak_2013,
       author = {{Cieplak}, Agnieszka M. and {Griest}, Kim},
        title = "{Improved Theoretical Predictions of Microlensing Rates for the Detection of Primordial Black Hole Dark Matter}",
      journal = {\apj},
         year = 2013,
        month = apr,
       volume = {767},
       number = {2},
          eid = {145},
        pages = {145},
          doi = {10.1088/0004-637X/767/2/145},
archivePrefix = {arXiv},
       eprint = {1210.7729},
 primaryClass = {astro-ph.CO}}

@ARTICLE{Aladin,
       author = {{Bonnarel}, F. and {Fernique}, P. and {Bienaym{\'e}}, O. and {Egret}, D. and {Genova}, F. and {Louys}, M. and {Ochsenbein}, F. and {Wenger}, M. and {Bartlett}, J.~G.},
        title = "{The ALADIN interactive sky atlas. A reference tool for identification of astronomical sources}",
      journal = {\aaps},
         year = 2000,
        month = apr,
       volume = {143},
        pages = {33-40},
          doi = {10.1051/aas:2000331}}

@misc{Decam_filter,
Author = {NOIRLab},
Title = {DECam Filters},
url={https://noirlab.edu/science/programs/ctio/filters/Dark-Energy-Camera/VR-filter},
Year = {2014}}

@article{Nidever_2021,
   title={The Second Data Release of the Survey of the MAgellanic Stellar History (SMASH)},
   volume={161},
   ISSN={1538-3881},
   DOI={10.3847/1538-3881/abceb7},
   number={2},
   journal={The Astronomical Journal},
   publisher={American Astronomical Society},
   author={Nidever, David L. and Olsen, Knut and Choi, Yumi and Ruiz-Lara, Tomas and Miller, Amy E. and Johnson, L. Clifton and Bell, Cameron P. M. and Blum, Robert D. and Cioni, Maria-Rosa L. and Gallart, Carme and Majewski, Steven R. and Martin, Nicolas F. and Massana, Pol and Monachesi, Antonela and Noël, Noelia E. D. and Sakowska, Joanna D. and van der Marel, Roeland P. and Walker, Alistair R. and Zaritsky, Dennis and Bell, Eric F. and Conn, Blair C. and de Boer, Thomas J. L. and Gruendl, Robert A. and Monelli, Matteo and Muñoz, Ricardo R. and Saha, Abhijit and Vivas, A. Katherina and Bernard, Edouard and Besla, Gurtina and Carballo-Bello, Julio A. and Dorta, Antonio and Martinez-Delgado, David and Goater, Alex and Rusakov, Vadim and Stringfellow, Guy S.},
   year={2021},
   month=jan, pages={74} }

@article{Dophot_1993,
    author = "Schechter, Paul L. and Mateo, Mario and Saha, Abhijit",
    title = "{DOPHOT, a CCD photometry program: Description and tests}",
    doi = "10.1086/133316",
    journal = "Publ. Astron. Soc. Pac.",
    volume = "105",
    pages = "1342",
    year = "1993"}

@article{Saha_2010,
   title={FIRST RESULTS FROM THE NOAO SURVEY OF THE OUTER LIMITS OF THE MAGELLANIC CLOUDS},
   volume={140},
   ISSN={1538-3881},
   DOI={10.1088/0004-6256/140/6/1719},
   number={6},
   journal={The Astronomical Journal},
   publisher={American Astronomical Society},
   author={Saha, Abhijit and Olszewski, Edward W. and Brondel, Brian and Olsen, Knut and Knezek, Patricia and Harris, Jason and Smith, Chris and Subramaniam, Annapurni and Claver, Jennifer and Rest, Armin and Seitzer, Patrick and Cook, Kem H. and Minniti, Dante and Suntzeff, Nicholas B.},
   year={2010},
   month=oct, pages={1719–1738} }

@article{Godines_2019,
title = {A machine learning classifier for microlensing in wide-field surveys},
journal = {Astronomy and Computing},
volume = {28},
pages = {100298},
year = {2019},
issn = {2213-1337},
doi = {https://doi.org/10.1016/j.ascom.2019.100298},
author = {D. Godines and E. Bachelet and G. Narayan and R.A. Street}}

@article{Husseiniova_2021,
    author = {Husseiniova, Andrea and McGill, Peter and Smith, Leigh C and Evans, N Wyn},
    title = "{A microlensing search of 700 million VVV light curves}",
    journal = {Monthly Notices of the Royal Astronomical Society},
    volume = {506},
    number = {2},
    pages = {2482-2502},
    year = {2021},
    month = {07},
    issn = {0035-8711},
    doi = {10.1093/mnras/stab1882},
}

@ARTICLE{PriceWhelan_2014,
       author = {{Price-Whelan}, Adrian M. and {Ag{\"u}eros}, Marcel A. and {Fournier}, Amanda P. and {Street}, Rachel and {Ofek}, Eran O. and {Covey}, Kevin R. and {Levitan}, David and {Laher}, Russ R. and {Sesar}, Branimir and {Surace}, Jason},
        title = "{Statistical Searches for Microlensing Events in Large, Non-uniformly Sampled Time-Domain Surveys: A Test Using Palomar Transient Factory Data}",
      journal = {\apj},
         year = 2014,
        month = jan,
       volume = {781},
       number = {1},
          eid = {35},
        pages = {35},
          doi = {10.1088/0004-637X/781/1/35}}

@article{Kim_2014,
   title={The EPOCH Project: I. Periodic variable stars in the EROS-2 LMC database⋆},
   volume={566},
   ISSN={1432-0746},
   url={http://dx.doi.org/10.1051/0004-6361/201323252},
   DOI={10.1051/0004-6361/201323252},
   journal={Astronomy &amp; Astrophysics},
   publisher={EDP Sciences},
   author={Kim, Dae-Won and Protopapas, Pavlos and Bailer-Jones, Coryn A. L. and Byun, Yong-Ik and Chang, Seo-Won and Marquette, Jean-Baptiste and Shin, Min-Su},
   year={2014},
   month=June, pages={A43} }

@article{Press_1969,
 ISSN = {00034851, 21688990},
 author = {S. James Press},
 journal = {The Annals of Mathematical Statistics},
 number = {1},
 pages = {188--196},
 publisher = {Institute of Mathematical Statistics},
 title = {On Serial Correlation},
 urldate = {2024-10-20},
 volume = {40},
 year = {1969}}

@article{VN1_1941,
 ISSN = {00034851},
 author = {J. von Neumann and R. H. Kent and H. R. Bellinson and B. I. Hart},
 journal = {The Annals of Mathematical Statistics},
 number = {2},
 pages = {153--162},
 publisher = {Institute of Mathematical Statistics},
 title = {The Mean Square Successive Difference},
 urldate = {2024-10-21},
 volume = {12},
 year = {1941}}

@article{VN2_1941,
 ISSN = {00034851},
 author = {John von Neumann},
 journal = {The Annals of Mathematical Statistics},
 number = {4},
 pages = {367--395},
 publisher = {Institute of Mathematical Statistics},
 title = {Distribution of the Ratio of the Mean Square Successive Difference to the Variance},
 urldate = {2024-10-21},
 volume = {12},
 year = {1941}}

@ARTICLE{Irwin_2007,
       author = {{Irwin}, Jonathan and {Irwin}, Mike and {Aigrain}, Suzanne and {Hodgkin}, Simon and {Hebb}, Leslie and {Moraux}, Estelle},
        title = "{The Monitor project: data processing and light curve production}",
      journal = {\mnras},
         year = 2007,
        month = mar,
       volume = {375},
       number = {4},
        pages = {1449-1462},
          doi = {10.1111/j.1365-2966.2006.11408.x}}

@article{Garcia_2010,
title = {Robust smoothing of gridded data in one and higher dimensions with missing values},
journal = {Computational Statistics & Data Analysis},
volume = {54},
number = {4},
pages = {1167-1178},
year = {2010},
issn = {0167-9473},
doi = {https://doi.org/10.1016/j.csda.2009.09.020},
author = {Damien Garcia}}

@article{Li_2019,
   title={Kepler Data Validation II–Transit Model Fitting and Multiple-planet Search},
   volume={131},
   ISSN={1538-3873},
   DOI={10.1088/1538-3873/aaf44d},
   number={996},
   journal={Publications of the Astronomical Society of the Pacific},
   publisher={IOP Publishing},
   author={Li, Jie and Tenenbaum, Peter and Twicken, Joseph D. and Burke, Christopher J. and Jenkins, Jon M. and Quintana, Elisa V. and Rowe, Jason F. and Seader, Shawn E.},
   year={2019},
   month=jan, pages={024506} }

@article{Thompson_2018,
   title={Planetary Candidates Observed by Kepler. VIII. A Fully Automated Catalog with Measured Completeness and Reliability Based on Data Release 25},
   volume={235},
   ISSN={1538-4365},
   DOI={10.3847/1538-4365/aab4f9},
   number={2},
   journal={The Astrophysical Journal Supplement Series},
   publisher={American Astronomical Society},
   author={Thompson, Susan E. and Coughlin, Jeffrey L. and Hoffman, Kelsey and Mullally, Fergal and Christiansen, Jessie L. and Burke, Christopher J. and Bryson, Steve and Batalha, Natalie and Haas, Michael R. and Catanzarite, Joseph and Rowe, Jason F. and Barentsen, Geert and Caldwell, Douglas A. and Clarke, Bruce D. and Jenkins, Jon M. and Li, Jie and Latham, David W. and Lissauer, Jack J. and Mathur, Savita and Morris, Robert L. and Seader, Shawn E. and Smith, Jeffrey C. and Klaus, Todd C. and Twicken, Joseph D. and Van Cleve, Jeffrey E. and Wohler, Bill and Akeson, Rachel and Ciardi, David R. and Cochran, William D. and Henze, Christopher E. and Howell, Steve B. and Huber, Daniel and Prša, Andrej and Ramírez, Solange V. and Morton, Timothy D. and Barclay, Thomas and Campbell, Jennifer R. and Chaplin, William J. and Charbonneau, David and Christensen-Dalsgaard, Jørgen and Dotson, Jessie L. and Doyle, Laurance and Dunham, Edward W. and Dupree, Andrea K. and Ford, Eric B. and Geary, John C. and Girouard, Forrest R. and Isaacson, Howard and Kjeldsen, Hans and Quintana, Elisa V. and Ragozzine, Darin and Shabram, Megan and Shporer, Avi and Aguirre, Victor Silva and Steffen, Jason H. and Still, Martin and Tenenbaum, Peter and Welsh, William F. and Wolfgang, Angie and Zamudio, Khadeejah A and Koch, David G. and Borucki, William J.},
   year={2018},
   month=apr, pages={38} }

@article{Lomb_1976,
  author = {N. R. Lomb},
  title = {Least-squares frequency analysis of unequally spaced data},
  journal = {Astrophysics and Space Science},
  volume = {39},
  number = {2},
  pages = {447--462},
  year = {1976},
  publisher = {Springer},
}

@article{Scargle_1983,
author = {Scargle, Jeffrey},
year = {1983},
month = {01},
pages = {},
title = {Studies in astronomical time series analysis. II - Statistical aspects of spectral analysis of unevenly spaced data},
volume = {263},
journal = {The Astrophysical Journal},
doi = {10.1086/160554}}

@article{GaiaVarInfo,
   title={GaiaData Release 3: Summary of the variability processing and analysis},
   volume={674},
   ISSN={1432-0746},
   DOI={10.1051/0004-6361/202244242},
   journal={Astronomy \& Astrophysics},
   publisher={EDP Sciences},
   author={Eyer, L. and Audard, M. and Holl, B. and Rimoldini, L. and Carnerero, M. I. and Clementini, G. and De Ridder, J. and Distefano, E. and Evans, D. W. and Gavras, P. and Gomel, R. and Lebzelter, T. and Marton, G. and Mowlavi, N. and Panahi, A. and Ripepi, V. and Wyrzykowski, Ł. and Nienartowicz, K. and Jevardat de Fombelle, G. and Lecoeur-Taibi, I. and Rohrbasser, L. and Riello, M. and García-Lario, P. and Lanzafame, A. C. and Mazeh, T. and Raiteri, C. M. and Zucker, S. and Ábrahám, P. and Aerts, C. and Aguado, J. J. and Anderson, R. I. and Bashi, D. and Binnenfeld, A. and Faigler, S. and Garofalo, A. and Karbevska, L. and Kóspál, Á and Kruszyńska, K. and Kun, M. and Lanza, A. F. and Leccia, S. and Marconi, M. and Messina, S. and Molinaro, R. and Molnár, L. and Muraveva, T. and Musella, I. and Nagy, Z. and Pagano, I. and Palaversa, L. and Plachy, E. and Prša, A. and Rybicki, K. A. and Shahaf, S. and Szabados, L. and Szegedi-Elek, E. and Trabucchi, M. and Barblan, F. and Grenon, M. and Roelens, M. and Süveges, M.},
   year={2023},
   month=jun, pages={A13} }

@article{Mulensmodel,
   title={Modeling microlensing events with MulensModel},
   volume={26},
   ISSN={2213-1337},
   DOI={10.1016/j.ascom.2018.11.001},
   journal={Astronomy and Computing},
   publisher={Elsevier BV},
   author={Poleski, R. and Yee, J.C.},
   year={2019},
   month=jan, pages={35–49} }

@article{Calcino_2018,
   title={Updating the MACHO fraction of the Milky Way dark halo with improved mass models},
   volume={479},
   ISSN={1365-2966},
   DOI={10.1093/mnras/sty1368},
   number={3},
   journal={Monthly Notices of the Royal Astronomical Society},
   publisher={Oxford University Press (OUP)},
   author={Calcino, Josh and García-Bellido, Juan and Davis, Tamara M},
   year={2018},
   month=may, pages={2889–2905} }

@article{Navarro_1997,
   title={A Universal Density Profile from Hierarchical Clustering},
   volume={490},
   ISSN={1538-4357},
   url={http://dx.doi.org/10.1086/304888},
   DOI={10.1086/304888},
   number={2},
   journal={The Astrophysical Journal},
   publisher={American Astronomical Society},
   author={Navarro, Julio F. and Frenk, Carlos S. and White, Simon D. M.},
   year={1997},
   month=dec, pages={493–508} }

@article{Blaineau_2020,
   title={Parallax in microlensing toward the Magellanic Clouds: Effect on detection efficiency and detectability},
   volume={636},
   ISSN={1432-0746},
   DOI={10.1051/0004-6361/202038005},
   journal={Astronomy \& Astrophysics},
   publisher={EDP Sciences},
   author={Blaineau, T. and Moniez, M.},
   year={2020},
   month=apr, pages={L9} }

@article{Battaglia_2005,
   title={The radial velocity dispersion profile of the Galactic halo: constraining the density profile of the dark halo of the Milky Way},
   volume={364},
   ISSN={1365-2966},
   DOI={10.1111/j.1365-2966.2005.09367.x},
   number={2},
   journal={Monthly Notices of the Royal Astronomical Society},
   publisher={Oxford University Press (OUP)},
   author={Battaglia, Giuseppina and Helmi, Amina and Morrison, Heather and Harding, Paul and Olszewski, Edward W. and Mateo, Mario and Freeman, Kenneth C. and Norris, John and Shectman, Stephen A.},
   year={2005},
   month=dec, pages={433–442} }

@article{Sajadian_2021,
   title={On the detection of free-floating planets through microlensing towards the Magellanic Clouds},
   volume={506},
   ISSN={1365-2966},
   DOI={10.1093/mnras/stab1907},
   number={3},
   journal={Monthly Notices of the Royal Astronomical Society},
   publisher={Oxford University Press (OUP)},
   author={Sajadian, Sedighe},
   year={2021},
   month=jul, pages={3615–3628} }

@article{Gyuk_2000,
   title={Self‐lensing Models of the Large Magellanic Cloud},
   volume={535},
   ISSN={1538-4357},
   DOI={10.1086/308834},
   number={1},
   journal={The Astrophysical Journal},
   publisher={American Astronomical Society},
   author={Gyuk, G. and Dalal, N. and Griest, K.},
   year={2000},
   month=may, pages={90–103} }

@article{Weinberg_2001,
   title={Structure of the Large Magellanic Cloud from 2MASS},
   volume={548},
   ISSN={1538-4357},
   DOI={10.1086/319001},
   number={2},
   journal={The Astrophysical Journal},
   publisher={American Astronomical Society},
   author={Weinberg, Martin D. and Nikolaev, Sergei},
   year={2001},
   month=feb, pages={712–726} }

@ARTICLE{Marel_2002,
       author = {{van der Marel}, Roeland P. and {Alves}, David R. and {Hardy}, Eduardo and {Suntzeff}, Nicholas B.},
        title = "{New Understanding of Large Magellanic Cloud Structure, Dynamics, and Orbit from Carbon Star Kinematics}",
      journal = {\aj},
         year = 2002,
        month = nov,
       volume = {124},
       number = {5},
        pages = {2639-2663},
          doi = {10.1086/343775}}

@ARTICLE{Nidever_2019,
       author = {{Nidever}, David L. and {Olsen}, Knut and {Choi}, Yumi and {de Boer}, Thomas J.~L. and {Blum}, Robert D. and {Bell}, Eric F. and {Zaritsky}, Dennis and {Martin}, Nicolas F. and {Saha}, Abhijit and {Conn}, Blair C. and {Besla}, Gurtina and {van der Marel}, Roeland P. and {No{\"e}l}, Noelia E.~D. and {Monachesi}, Antonela and {Stringfellow}, Guy S. and {Massana}, Pol and {Cioni}, Maria-Rosa L. and {Gallart}, Carme and {Monelli}, Matteo and {Martinez-Delgado}, David and {Mu{\~n}oz}, Ricardo R. and {Majewski}, Steven R. and {Vivas}, A. Katherina and {Walker}, Alistair R. and {Kaleida}, Catherine and {Chu}, You-Hua},
        title = "{Exploring the Very Extended Low-surface-brightness Stellar Populations of the Large Magellanic Cloud with SMASH}",
      journal = {\apj},
         year = 2019,
        month = apr,
       volume = {874},
       number = {2},
          eid = {118},
        pages = {118},
          doi = {10.3847/1538-4357/aafaf7}}

@article{Kallivayalil_2013,
   title={THIRD-EPOCH MAGELLANIC CLOUD PROPER MOTIONS. I.HUBBLE SPACE TELESCOPE/WFC3 DATA AND ORBIT IMPLICATIONS},
   volume={764},
   ISSN={1538-4357},
   DOI={10.1088/0004-637x/764/2/161},
   number={2},
   journal={The Astrophysical Journal},
   publisher={American Astronomical Society},
   author={Kallivayalil, Nitya and van der Marel, Roeland P. and Besla, Gurtina and Anderson, Jay and Alcock, Charles},
   year={2013},
   month=feb, pages={161} }

@article{Alves_2004,
   title={The Stellar Halo in the Large Magellanic Cloud: Mass, Luminosity, and Microlensing Predictions},
   volume={601},
   ISSN={1538-4357},
   DOI={10.1086/382130},
   number={2},
   journal={The Astrophysical Journal},
   publisher={American Astronomical Society},
   author={Alves, David R.},
   year={2004},
   month=jan, pages={L151–L154} }

@ARTICLE{Pietrzyski_2019,
       author = {{Pietrzy{\'n}ski}, G. and {Graczyk}, D. and {Gallenne}, A. and {Gieren}, W. and {Thompson}, I.~B. and {Pilecki}, B. and {Karczmarek}, P. and {G{\'o}rski}, M. and {Suchomska}, K. and {Taormina}, M. and {Zgirski}, B. and {Wielg{\'o}rski}, P. and {Ko{\l}aczkowski}, Z. and {Konorski}, P. and {Villanova}, S. and {Nardetto}, N. and {Kervella}, P. and {Bresolin}, F. and {Kudritzki}, R.~P. and {Storm}, J. and {Smolec}, R. and {Narloch}, W.},
        title = "{A distance to the Large Magellanic Cloud that is precise to one per cent}",
      journal = {\nat},
         year = 2019,
        month = mar,
       volume = {567},
       number = {7747},
        pages = {200-203},
          doi = {10.1038/s41586-019-0999-4}}

@article{Smyth_2020,
   title={Updated constraints on asteroid-mass primordial black holes as dark matter},
   volume={101},
   ISSN={2470-0029},
   DOI={10.1103/physrevd.101.063005},
   number={6},
   journal={Physical Review D},
   publisher={American Physical Society (APS)},
   author={Smyth, Nolan and Profumo, Stefano and English, Samuel and Jeltema, Tesla and McKinnon, Kevin and Guhathakurta, Puragra},
   year={2020},
   month=mar }

@ARTICLE{dotter_MIST,
       author = {{Dotter}, Aaron},
        title = "{MESA Isochrones and Stellar Tracks (MIST) 0: Methods for the Construction of Stellar Isochrones}",
      journal = {\apjs},
         year = 2016,
        month = jan,
       volume = {222},
       number = {1},
          eid = {8},
        pages = {8},
          doi = {10.3847/0067-0049/222/1/8}}

@ARTICLE{choi_MIST,
       author = {{Choi}, Jieun and {Dotter}, Aaron and {Conroy}, Charlie and {Cantiello}, Matteo and {Paxton}, Bill and {Johnson}, Benjamin D.},
        title = "{Mesa Isochrones and Stellar Tracks (MIST). I. Solar-scaled Models}",
      journal = {\apj},
         year = 2016,
        month = jun,
       volume = {823},
       number = {2},
          eid = {102},
        pages = {102},
          doi = {10.3847/0004-637X/823/2/102}}

@ARTICLE{Paxton_MISTA,
       author = {{Paxton}, Bill and {Bildsten}, Lars and {Dotter}, Aaron and {Herwig}, Falk and {Lesaffre}, Pierre and {Timmes}, Frank},
        title = "{Modules for Experiments in Stellar Astrophysics (MESA)}",
      journal = {\apjs},
         year = 2011,
        month = jan,
       volume = {192},
       number = {1},
          eid = {3},
        pages = {3},
          doi = {10.1088/0067-0049/192/1/3}}

@ARTICLE{Paxon_MISTB,
       author = {{Paxton}, Bill and {Cantiello}, Matteo and {Arras}, Phil and {Bildsten}, Lars and {Brown}, Edward F. and {Dotter}, Aaron and {Mankovich}, Christopher and {Montgomery}, M.~H. and {Stello}, Dennis and {Timmes}, F.~X. and {Townsend}, Richard},
        title = "{Modules for Experiments in Stellar Astrophysics (MESA): Planets, Oscillations, Rotation, and Massive Stars}",
      journal = {\apjs},
         year = 2013,
        month = sep,
       volume = {208},
       number = {1},
          eid = {4},
        pages = {4},
          doi = {10.1088/0067-0049/208/1/4}}

@ARTICLE{Paxon_MISTC,
       author = {{Paxton}, Bill and {Marchant}, Pablo and {Schwab}, Josiah and {Bauer}, Evan B. and {Bildsten}, Lars and {Cantiello}, Matteo and {Dessart}, Luc and {Farmer}, R. and {Hu}, H. and {Langer}, N. and {Townsend}, R.~H.~D. and {Townsley}, Dean M. and {Timmes}, F.~X.},
        title = "{Modules for Experiments in Stellar Astrophysics (MESA): Binaries, Pulsations, and Explosions}",
      journal = {\apjs},
         year = 2015,
        month = sep,
       volume = {220},
       number = {1},
          eid = {15},
        pages = {15},
          doi = {10.1088/0067-0049/220/1/15}}

@article{Narloch_2022,
   title={Metallicities and ages for star clusters and their surrounding fields in the Large Magellanic Cloud},
   volume={666},
   ISSN={1432-0746},
   DOI={10.1051/0004-6361/202243378},
   journal={Astronomy & Astrophysics},
   publisher={EDP Sciences},
   author={Narloch, W. and Pietrzyński, G. and Gieren, W. and Piatti, A. E. and Karczmarek, P. and Górski, M. and Graczyk, D. and Smolec, R. and Hajdu, G. and Suchomska, K. and Zgirski, B. and Wielgórski, P. and Pilecki, B. and Taormina, M. and Kałuszyński, M. and Pych, W. and Rojas García, G. and Lewis, M. O.},
   year={2022},
   month=oct, pages={A80} }

@article{Sugiyama_2020,
	doi = {10.1093/mnras/staa407},
	year = 2020,
	month = {feb},
	publisher = {Oxford University Press ({OUP})},
	volume = {493},
	number = {3},
	pages = {3632--3641},
	author = {Sunao Sugiyama and Toshiki Kurita and Masahiro Takada},
	title = {On the wave optics effect on primordial black hole constraints from optical microlensing search},
	journal = {Monthly Notices of the Royal Astronomical Society}}

@ARTICLE{Nakamura_1999,
       author = {{Nakamura}, T.~T. and {Deguchi}, S.},
        title = "{Wave Optics in Gravitational Lensing}",
      journal = {Progress of Theoretical Physics Supplement},
         year = 1999,
        month = jan,
       volume = {133},
        pages = {137-153},
          doi = {10.1143/PTPS.133.137}}

@article{Matsunaga_2006,
   title={The finite source size effect and wave optics in gravitational lensing},
   volume={2006},
   ISSN={1475-7516},
   DOI={10.1088/1475-7516/2006/01/023},
   number={01},
   journal={Journal of Cosmology and Astroparticle Physics},
   publisher={IOP Publishing},
   author={Matsunaga, Norihito and Yamamoto, Kazuhiro},
   year={2006},
   month=jan, pages={023–023} }

@article{Tisserand_2007,
   title={Limits on the Macho content of the Galactic Halo from the EROS-2 Survey of the Magellanic Clouds},
   volume={469},
   ISSN={1432-0746},
   url={http://dx.doi.org/10.1051/0004-6361:20066017},
   DOI={10.1051/0004-6361:20066017},
   number={2},
   journal={Astronomy & Astrophysics},
   publisher={EDP Sciences},
   author={Tisserand, P. and Le Guillou, L. and Afonso, C. and Albert, J. N. and Andersen, J. and Ansari, R. and Aubourg, É. and Bareyre, P. and Beaulieu, J. P. and Charlot, X. and Coutures, C. and Ferlet, R. and Fouqué, P. and Glicenstein, J. F. and Goldman, B. and Gould, A. and Graff, D. and Gros, M. and Haissinski, J. and Hamadache, C. and de Kat, J. and Lasserre, T. and Lesquoy, É. and Loup, C. and Magneville, C. and Marquette, J. B. and Maurice, É. and Maury, A. and Milsztajn, A. and Moniez, M. and Palanque-Delabrouille, N. and Perdereau, O. and Rahal, Y. R. and Rich, J. and Spiro, M. and Vidal-Madjar, A. and Vigroux, L.},
   year={2007},
   month=apr, pages={387–404} }

@article{Griest_2014,
   title={EXPERIMENTAL LIMITS ON PRIMORDIAL BLACK HOLE DARK MATTER FROM THE FIRST 2 YR OFKEPLERDATA},
   volume={786},
   ISSN={1538-4357},
   url={http://dx.doi.org/10.1088/0004-637X/786/2/158},
   DOI={10.1088/0004-637x/786/2/158},
   number={2},
   journal={The Astrophysical Journal},
   publisher={American Astronomical Society},
   author={Griest, Kim and Cieplak, Agnieszka M. and Lehner, Matthew J.},
   year={2014},
   month=apr, pages={158} }

@article{Gorton_2022,
   title={Effect of clustering on primordial black hole microlensing constraints},
   volume={2022},
   ISSN={1475-7516},
   url={http://dx.doi.org/10.1088/1475-7516/2022/08/035},
   DOI={10.1088/1475-7516/2022/08/035},
   number={08},
   journal={Journal of Cosmology and Astroparticle Physics},
   publisher={IOP Publishing},
   author={Gorton, Matthew and Green, Anne M.},
   year={2022},
   month=aug, pages={035} }

@article{Calchinovati_2013,
    author = {Calchi Novati, S. and Mirzoyan, S. and Jetzer, Ph. and Scarpetta, G.},
    title = {Microlensing towards the SMC: a new analysis of OGLE and EROS results},
    journal = {Monthly Notices of the Royal Astronomical Society},
    volume = {435},
    number = {2},
    pages = {1582-1597},
    year = {2013},
    month = {08},
    issn = {0035-8711},
    doi = {10.1093/mnras/stt1402},
    url = {https://doi.org/10.1093/mnras/stt1402},
    eprint = {https://academic.oup.com/mnras/article-pdf/435/2/1582/3533110/stt1402.pdf},
}

@article{Feldman_1998,
   title={Unified approach to the classical statistical analysis of small signals},
   volume={57},
   ISSN={1089-4918},
   url={http://dx.doi.org/10.1103/PhysRevD.57.3873},
   DOI={10.1103/physrevd.57.3873},
   number={7},
   journal={Physical Review D},
   publisher={American Physical Society (APS)},
   author={Feldman, Gary J. and Cousins, Robert D.},
   year={1998},
   month=apr, pages={3873–3889} }

@misc{PBHBounds,
  author = {{Bradley J. Kavanagh}},
  title = {bradkav/PBHbounds: Release version 0.1},
  howpublished = {https://doi.org/10.5281/zenodo.3538999},
  year = {2019},
}

\bsp	
\label{lastpage}
\end{document}